\documentclass[11pt]{article}

\usepackage[final]{acl}

\usepackage{times}
\usepackage{latexsym}

\usepackage[T1]{fontenc}

\usepackage[utf8]{inputenc}

\usepackage{enumitem}
\usepackage{microtype}

\usepackage{inconsolata}

\usepackage{graphicx}
\usepackage{amsmath}
\usepackage{amssymb}
\usepackage{placeins}
\usepackage{multirow}
\usepackage{amssymb}
\usepackage{tabularx}
\usepackage{booktabs}
\usepackage{amsmath}
\DeclareMathOperator*{\argmax}{arg\,max}
\usepackage{verbatim}
\usepackage[table]{xcolor}
\usepackage{alphalph}

\title{``Not in My Backyard'': LLMs Uncover Online and Offline Social Biases Against Homelessness}

\author{
  Jonathan A. Karr Jr.$^{1}$
  \and Benjamin F. Herbst$^{1}$
  \and Matthew L. Sisk$^{1}$
  \and Xueyun Li$^{1}$
  \AND
  Ting Hua$^{1}$
  \and Matthew Hauenstein$^{1}$
  \and Georgina Curto$^{2}$
  \and Nitesh V. Chawla$^{1}$ \\[0.5em]
  $^{1}$University of Notre Dame, USA \\
  $^{2}$United Nations University Institute in Macau, Macau SAR, China \\[0.25em]
  \texttt{\{jkarr, bherbst, msisk1, xli44, thua, mhauenst, nchawla\}@nd.edu},
  \texttt{curto@unu.edu}
}

\begin{document}
\maketitle
\begin{abstract}
Homelessness is a persistent social challenge, impacting millions worldwide. Over 876,000 people experiencing homelessness (PEH) were recorded in the U.S.\ in 2025. Social bias is a significant barrier to alleviating homelessness, shaping public perception and influencing policymaking. Because online textual media and offline city council discourse both reflect and influence public opinion, they provide valuable signals for identifying and tracking social biases against PEH. We release the first multi-domain PEH bias corpus with a 16-category multi-label taxonomy: a 1,698-item stratified gold-standard set annotated by partner-trained raters, plus 48,389 GPT-4.1-labeled texts, drawn from Reddit, X (formerly Twitter), news, and council meeting transcripts across ten U.S.\ cities (2015-2025). We benchmark six prompted LLMs on the gold-standard set and complement F1 with prevalence-gap audits. Moderate F1 coexists with large miscalibration: every model over-tags ``not in my backyard'' (NIMBY) ($+11.5$ pp) and under-detects factual claims ($-30.5$ pp). Error analysis on consensus false positives reveals that models treat housing vocabulary and question form as opposition proxies, producing NIMBY false positives on pro-service text. The corpus and audit protocol support municipal PEH stigma monitoring without treating teacher labels as ground truth.\end{abstract}
\noindent Content Warning: This paper presents textual examples that may be offensive or upsetting.\\
\href{https://github.com/Homelessness-Project/Multidomain-PEH-Classification}{Code: https://github.com/Homelessness-Project/Multidomain-PEH-Classification}\\
\href{https://zenodo.org/records/16877412}{Dataset:https://zenodo.org/records/16877412}

\section{Introduction}

Homelessness is a persistent social challenge: the Organization for Economic Co-operation and Development (OECD) reports 2.2 million people experiencing homelessness (PEH) across OECD and European Union countries \citep{oecd2024homelessness}, and more than 876{,}000 were recorded in the U.S.\ in 2025, the highest count on record \citep{hud2025no25-153}. Stigma against PEH dehumanizes those affected and makes it harder for policymakers to approve shelters and services \citep{curto2024crime,curto2025tackling,clifford2017explaining}. Public opinion in online media and city council meetings offers a relatively affordable signal of how communities talk about homelessness \citep{AlanChan2021, Mislove2011}.

Yet partners monitoring shelter siting need more than overlap scores: they need to know whether an automated monitor \emph{inflates or deflates} stigma categories that shape public backlash. No existing resource provides multi-domain, human-audited PEH bias labels together with a calibration audit for that deployment question. Fine-grained multi-label classification is necessary but not sufficient for reliable stigma monitoring.
\begin{figure}
    \centering
    \includegraphics[width=1\linewidth]{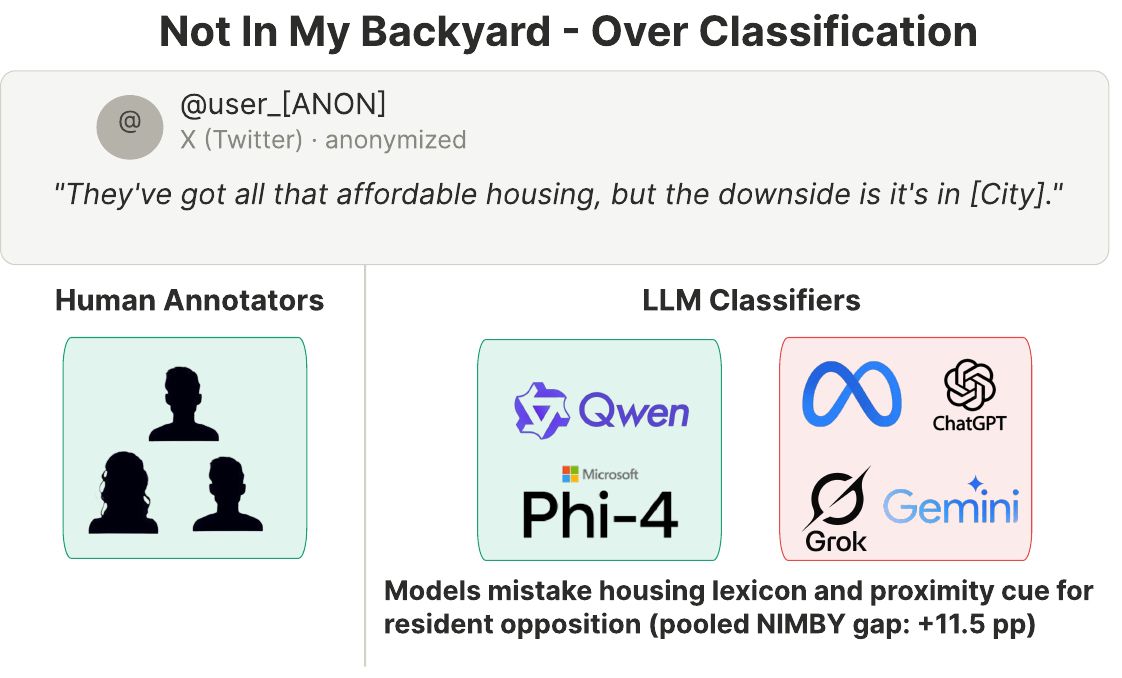}
    \caption{Four of six zero-shot LLMs tag \textit{not in my backyard (NIMBY)} on an X post that gold annotators labeled only as \textit{provide a fact or claim} and \textit{express own opinion} (GPT 4.1, Gemini 2.5 Pro, Grok 4, LLaMA 3.2 3B; Phi-4 Mini and Qwen 2.5 7B do not): ``They've got all that affordable housing, but the downside is it's in [City].''}
    \label{fig:nimby_fp}
\end{figure}
Prior PEH bias work (e.g., OATH \citep{ranjit2024oath}) releases expert reference annotations on X but uses a single \textit{homelessness} keyword and does not audit label-prevalence shift when scaling with LLMs. We address these gaps with the following contributions:
\begin{itemize}[nosep,leftmargin=*]
\item We release a multi-domain PEH bias corpus of 50,087 texts from Reddit, X, news, and city council meeting transcripts across ten U.S.\ cities (2015-2025), collected with an 11-term PEH lexicon (Appendix~\ref{sec:peh_lexicon}).
\item We construct a 1,698-item stratified gold-standard set with a 16-category multi-label taxonomy, annotated with the support of domain experts in non-profit organizations [A] and [B] (Appendix~\ref{sec:stakeholder_collaboration}).
\item We propose a prevalence-gap audit protocol that complements F1 with systematic calibration checks on the gold standard.
\item We benchmark six prompted LLMs (zero-/few-shot) and provide error analysis showing that models use housing vocabulary and question form as NIMBY proxies.
\end{itemize}

\section{Corpus and Annotation}
\label{sec:corpus}

We collected discourse from ten U.S.\ cities in two kNN-matched clusters (five smaller cities, five larger cities), matched on population, homelessness rate, and Gini index. Cities required $\geq$50 Reddit PEH posts (2015-2025) for inclusion. For each city, we retrieved Reddit, X, and news text using an 11-term PEH lexicon (Appendix~\ref{sec:peh_lexicon}) and transcribed council meetings. Council-meeting data collection varied by municipality: San Francisco publishes verbatim transcripts; seven cities provide audio/video recordings that we transcribed with Whisper; Kalamazoo and Baltimore publish only short agenda summaries, so we excluded them from transcript-based council analyses while retaining their online and news data. We anonymized all text with spaCy \citep{HonnibalMontani2020}, drew a stratified sample of 1,698 items for partner annotation, and labeled the remaining corpus with GPT-4.1 few-shot prompts (Appendix Figure~\ref{fig:pipeline}). Table~\ref{tab:summary_data} summarizes 50,087 texts (anonymization details in Appendix~\ref{sec:anonymization_details}). For X, we inferred user locations with a multi-stage geolocation pipeline (Appendix~\ref{sec:geolocation}) because native GPS geolocation is available for $<$1\% of collected X posts (81/16{,}472); the pipeline locates $\approx$80\% of users in our 2025 subsample.

\begin{table*}[htpb!]
  \centering
  \small
  \begin{tabularx}{\textwidth}{l *{10}{>{\centering\arraybackslash}X}}
  \toprule
  City & \multicolumn{2}{c}{Reddit} & \multicolumn{2}{c}{News} & \multicolumn{3}{c}{X (Twitter)} & \multicolumn{2}{c}{City Council} & \\
  & Posts & Comments & Articles & Paragraphs & Posts & Geolocated & Non-Reposts & Meetings & Comments & Grand Total \\
  \midrule
    \textbf{South Bend}& 62 & \textbf{196} & 28 & \textbf{49} & 96 & 5 & \textbf{60}  &86& \textbf{330} & \textbf{635}\\
    \textbf{Rockford}& 43 & \textbf{188} & 5 & \textbf{9} & 98 & 0 & \textbf{37}  &344& \textbf{243} & \textbf{477}\\
    \textbf{Kalamazoo}& 209 & \textbf{1885} & 6 & \textbf{11} & 99 & 1 & \textbf{32}  &N/A& N/A & \textbf{1928}\\
    \textbf{Scranton}& 13 & \textbf{79} & 74 & \textbf{159} & 92 & 2 & \textbf{51}  &431& \textbf{514} & \textbf{803}\\
    \textbf{Fayetteville}& 34 & \textbf{102} & 13 & \textbf{32} & 97 & 3 & \textbf{77}  &233& \textbf{1043} & \textbf{1254}\\
    \textbf{San Francisco}& 714 & \textbf{14777} & 738 & \textbf{1537} & 9168 & 23 & \textbf{2137}  &25& \textbf{14} & \textbf{18465}\\
    \textbf{Portland}& 751 & \textbf{15301} & 179 & \textbf{397} & 5574 & 38 & \textbf{1090}  &372& \textbf{6618} & \textbf{23406}\\
    \textbf{Buffalo}& 151 & \textbf{589} & 99 & \textbf{196} & 685 & 1 & \textbf{100}  &211& \textbf{135} & \textbf{1020}\\
    \textbf{Baltimore}& 246 & \textbf{1215} & 85 & \textbf{156} & 464 & 7 & \textbf{222}  &N/A& N/A & \textbf{1593}\\
    \textbf{El Paso}& 40 & \textbf{154} & 17 & \textbf{31} & 99 & 1 & \textbf{37}  &74& \textbf{284} & \textbf{506}\\
    \midrule
    \textbf{Grand Total} & 2263 & \textbf{34486} & 1244 & \textbf{2577}& 16472 & 81 & \textbf{3843}&1776 & \textbf{9181} & \textbf{50087} \\
  \bottomrule
  \end{tabularx}
  \caption{Dataset Summary}
  \label{tab:summary_data}
\end{table*}

\subsection{Taxonomy and Gold-Standard Reference Labels}
\label{sec:bias_classification}
We formalize PEH bias detection as multi-label classification: given text $x_i$, predict $\hat{y}_i \in \{0,1\}^{16}$. Posts routinely carry multiple simultaneous labels (e.g., \textit{express opinion}, \textit{harmful generalization}, and \textit{not in my backyard}, so we retain co-occurrence structure rather than collapsing to a single label.

We extend OATH's nine frames \citep{ranjit2024oath} with Reddit- and council-specific types (genuine vs.\ rhetorical questions; fact/claim vs.\ observation; attributed opinions; \textit{racist}) that X-centric frames omit. Domain experts from specialized non-profit organizations defined the five-category \textit{Negative Bias Frame} for shelter-siting monitoring: rhetorical question, NIMBY, harmful generalization, deserving/undeserving, and racist (definitions in Appendix~\ref{sec:classification_categories}; rationale in Appendix~\ref{sec:taxonomy_rationale}).

Following prior PEH framing work that releases \emph{expert-only} reference annotations for evaluation \citep{ranjit2024oath} and the practice of using manually annotated gold standards in low-resource NLP \citep{cardoso2014gold}, three annotators trained in collaboration with non-profit organizations [A] and [B] labeled 1,698 stratified items \citep{liberty2016stratified} (up to 50 per city $\times$ source). This held-out \textbf{gold-standard set} is our reference for model metrics; corpus-scale GPT-4.1 labels are teacher annotations, not ground truth. Mean inter-annotator per-cell unanimity is 78.38\%; disagreement concentrates on subjective categories (Cohen's $\kappa$ in Appendix~\ref{sec:annotator_agreement}). We use soft labeling \citep{fornaciari2021beyond}: for sample $i$ and category $j$,
\begin{equation}
y_{ij} = \mathbb{1}\left[\sum_{k=1}^{3} a_{ijk} \geq 2\right], \quad a_{ijk} \in \{0,1\}.
\end{equation}
We threshold at $\tau{=}0.5$ (majority vote) for binary evaluation and prevalence-gap reporting.

\section{LLM Evaluation on the Gold-Standard Set}
\label{sec:llm_eval}
We benchmark six prompted LLMs as multi-label classifiers against gold-standard reference labels only: GPT-4.1, Gemini 2.5 Pro, Grok-4, LLaMA 3.2 3B, Qwen 2.5 7B, and Phi-4 Mini (zero- and few-shot with a prompt in Appendix~\ref{sec:llm_prompt}). Local models are included for privacy-sensitive deployment; closed-source APIs serve as performance references.

The gold-standard set supports evaluation but is too small for fine-tuning; five-shot GPT-4.1 improves only marginally over zero-shot (41.57 $\rightarrow$ 43.45 macro-F1 average). Corpus-scale GPT-4.1 labels are applied only after auditing prevalence gaps on the gold standard (Section~\ref{sec:ro3_calibration}).

\section{Results}
\label{sec:results}

\subsection{LLM Benchmarks on the Gold-Standard Set}
\label{sec:llm_benchmarks}
Table~\ref{tab:llm_benchmark} reports macro-F1 (mean per-label F1 at $\tau{=}0.5$) by source for all six prompted LLMs (zero- and few-shot). Closed APIs reach $\approx$43 macro-F1 (few-shot family average); local models $\approx$32; few-shot helps marginally and does not close the gap. Gemini 2.5 Pro attains the highest few-shot average (44.47); GPT-4.1 is within $\approx$1 point (43.45) and is our reference teacher for exploratory corpus labeling (Appendix~\ref{sec:teacher_model}). At face value, moderate F1 suggests LLMs are viable PEH bias monitors. The next subsections show that overlap alone does not reveal systematic prevalence shift. Full micro-F1 grids are in Appendix~\ref{sec:llm_results}.

\begin{table*}[t]
\centering
\scriptsize
\setlength{\tabcolsep}{1.5pt}
\begin{tabular}{l *{12}{c}}
\toprule
 & \multicolumn{2}{c}{GPT-4.1} & \multicolumn{2}{c}{Gemini} & \multicolumn{2}{c}{Grok} & \multicolumn{2}{c}{LLaMA} & \multicolumn{2}{c}{Phi-4} & \multicolumn{2}{c}{Qwen} \\
\cmidrule(lr){2-3}\cmidrule(lr){4-5}\cmidrule(lr){6-7}\cmidrule(lr){8-9}\cmidrule(lr){10-11}\cmidrule(lr){12-13}
Source & Z & F & Z & F & Z & F & Z & F & Z & F & Z & F \\
\midrule
Reddit (Ma) & 50.57 & 51.28 & 50.90 & \textbf{51.52} & 38.51 & 38.61 & 19.42 & 29.66 & 15.33 & 24.09 & 35.62 & 38.24 \\
News (Ma) & 34.79 & 37.93 & 35.99 & \textbf{39.08} & 26.04 & 28.69 & 15.05 & 15.94 & 17.11 & 21.30 & 22.62 & 25.12 \\
Meeting (Ma) & 37.70 & 39.96 & 38.06 & \textbf{41.38} & 29.46 & 34.46 & 13.84 & 21.54 & 18.12 & 20.81 & 26.82 & 30.32 \\
X (Ma) & 43.21 & 44.65 & 45.06 & \textbf{45.91} & 26.57 & 35.37 & 16.57 & 25.50 & 11.90 & 26.60 & 28.00 & 35.45 \\
\midrule
Avg (Ma) & 41.57 & 43.45 & 42.50 & \textbf{44.47} & 30.14 & 34.28 & 16.22 & 23.16 & 15.62 & 23.20 & 28.27 & 32.28 \\
\bottomrule
\end{tabular}
\caption{Prompted LLMs on the gold-standard set (macro-F1; Z = zero-shot, F = few-shot). Bold: best few-shot score per source among the six models.}
\label{tab:llm_benchmark}
\end{table*}
\begin{table*}[!tbp]
\centering
\footnotesize
\setlength{\tabcolsep}{3pt}
\begin{tabular}{@{}lrrrrrrr@{}}
\toprule
\textbf{Category} & \textbf{LLaMA} & \textbf{Phi-4} & \textbf{Qwen} & \textbf{GPT-4.1} & \textbf{Gemini} & \textbf{Grok} & \textbf{Pooled} \\
\midrule
Not in my backyard & +15.9 & +14.4 & +27.4 & +4.0 & +4.8 & +2.4 & +11.5 \\
Provide fact/claim & $-$55.1 & $-$61.9 & $-$23.2 & $-$7.6 & $-$1.2 & $-$34.3 & $-$30.5 \\
Express opinion & +16.4 & $-$16.4 & +25.2 & +10.7 & +7.0 & $-$20.2 & +3.8 \\
Harmful generalization & +20.1 & 0.0 & +27.2 & +9.8 & +9.8 & 0.0 & +11.1 \\
\bottomrule
\end{tabular}
\caption{Prevalence gap (pp) by model on the gold-standard set ($\hat{\pi}_{\mathrm{model}}-\pi_{\mathrm{ref}}$; pooled zero- and few-shot, all sources). \textbf{Pooled} = six-model average. Positive = over-tagging.}
\label{tab:calibration_by_model}
\end{table*}
\subsection{Prevalence Gaps vs.\ Gold-Standard Reference Labels}
\label{sec:ro3_calibration}
A model can attain moderate F1 yet systematically shift prevalence. We report $\hat{\pi}_{\mathrm{model}}-\pi_{\mathrm{ref}}$ (percentage points), where $\pi_{\mathrm{ref}}$ is the positive rate on gold-standard soft labels at $\tau{=}0.5$, pooling zero- and few-shot predictions and all four sources per model. Gaps of 10-60\,pp change which narratives appear prevalent in monitoring.

Table~\ref{tab:calibration_by_model} shows per-model drift on NIMBY, \textit{provide fact/claim}, \textit{express opinion}, and \textit{harmful generalization}. Every model over-tags NIMBY; every model under-detects factual claims. The pattern is not uniform: \textbf{Qwen trails GPT-4.1 by $\approx$10 macro-F1 points but its NIMBY over-tagging is $\mathbf{7\times}$ worse} ($+27.4$ vs.\ $+4.0$ pp). Privacy-sensitive local deployments therefore need \emph{stricter} prevalence-gap auditing than headline F1 suggests. Pooled NIMBY gap $+11.5$ pp has bootstrap 95\% CI [10.6, 12.6] ($n{=}1{,}698$; Appendix~\ref{sec:prevalence_uncertainty}).

\textit{Provide fact/claim} under-detection makes communities look less evidence-based than they are; NIMBY over-tagging makes them look more opposed to shelters than they are. LLMs used as bias monitors can therefore \textbf{distort what they measure} even when F1 looks acceptable.

\subsection{Why Models Over-Tag NIMBY}
\label{sec:nimby_error_main}
We analyze consensus NIMBY false positives: gold-negative items where $\geq 3$ of six zero-shot models tag NIMBY ($n{=}117$; 6.9\% of the gold set). Table~\ref{tab:nimby_triggers_main} compares lexical trigger rates in consensus FPs, true negatives, and gold NIMBY positives (full cohort definitions and triggers in Appendix~\ref{sec:nimby_error_analysis}).

Consensus FPs are elevated for \textbf{housing/homeless vocabulary} (89.7\% vs.\ 72.2\% among true negatives; +17.6 pp) and \textbf{question marks} (32.5\% vs.\ 9.7\%; +22.8 pp), but \textbf{opposition verbs} appear at only 8.5\%, below the 16.7\% rate among gold NIMBY positives. Affordable-housing phrases are slightly \emph{less} common among consensus FPs than among true negatives ($-$5.7 pp). Models appear to use ``mentions housing'' and question form as proxies for ``opposes housing,'' rather than detecting actual resident opposition to siting.

\begin{table}[t]
\centering
\footnotesize
\setlength{\tabcolsep}{4pt}
\begin{tabular}{@{}lrrr@{}}
\toprule
\textbf{Trigger} & \textbf{FP (\%)} & \textbf{TN (\%)} & \textbf{Gold$+$ (\%)} \\
\midrule
Question mark & 32.5 & 9.7 & 31.0 \\
Housing / homeless lexicon & 89.7 & 72.2 & 92.9 \\
Opposition verbs & 8.5 & 4.0 & 16.7 \\
Affordable housing phrase & 11.1 & 16.8 & 11.9 \\
\bottomrule
\end{tabular}
\caption{Trigger rates in consensus NIMBY false positives (FP, $n{=}117$), true negatives (TN, $n{=}672$), and gold NIMBY positives ($n{=}42$). Appendix Table~\ref{tab:nimby_triggers} lists all cues.}
\label{tab:nimby_triggers_main}
\end{table}

Figure~\ref{fig:nimby_fp} shows one multi-model false positive: four of six tag NIMBY on X (gold: provide a fact/claim and express own opinion only). Two additional patterns appear. Consensus false positive ($\geq 3$ of six): five of six tag NIMBY on pro-housing Reddit text gold-labeled only as \textit{express own opinion}:``Everybody wants more affordable housing\ldots but nobody seems willing to abandon\ldots a two-car garage, backyard and white-picket fence.'' Additionally, Qwen tags NIMBY on ``Did you read the same article I did? He wants to grow the amount of affordable housing from 25\% to 40\%,'' which humans labeled \textit{provide a fact or claim} (with partial annotator agreement on \textit{solutions/interventions}), not NIMBY. Conversely, five of six omit \textit{provide fact/claim} on supportive council language that annotators coded as a claim: ``Those people are not throw away people just because they are homeless.'' Partners should pair F1 with recurring gold-standard prevalence-gap audits before acting on corpus dashboards (Appendix~\ref{sec:municipal_deployment}).

\subsection{Implications for Deployment}
\label{sec:deployment_implications}
Non-profit partners Motels4Now and Broadway Christian
Parish are preparing to utilize this in cities. Through monthly meetings, Marshall Smith co-designed the taxonomy, and an anti-NIMBYism outreach campaign will follow once prevalence-gap adjustments are in place. Intended use is \emph{directional} monitoring. We recommend pairing any rollout with recurring gold-standard audits and prevalence-gap caps (Appendix~\ref{sec:municipal_deployment}); LoRA-tuned encoders such as ModernBERT approach closed-source models on micro-F1 (Appendix~\ref{sec:evaluation_framing}), offering a potential low-cost local path if prevalence gaps are audited and capped on the gold-standard reference set before scaling to public dashboards.

\section{Conclusion}

 Policy for PEH depends on local community discussion, yet stigma monitoring has lacked a multi-domain NLP resource. We release the first corpus and gold-standard reference set spanning Reddit, X, news, and city council meetings across ten U.S.\ cities, co-designed with non-profit partners. Benchmarking six prompted LLMs against gold-standard labels reveals that moderate F1 coexists with systematic miscalibration: every model over-tags NIMBY ($+11.5$ pp pooled) and under-detects factual claims, making communities look more opposed to shelters and less evidence-based than they are. Error analysis traces these failures to shallow lexical proxies rather than genuine opposition signals, and local models magnify the problem with far larger prevalence gaps than closed APIs. \textbf{Prevalence-gap audits should therefore accompany F1} before any LLM-labeled dashboard is treated as ground truth. Together, the corpus, gold-standard reference set, and audit protocol provide municipalities with a reproducible foundation for monitoring PEH stigma, one that foregrounds calibration alongside accuracy.


\section*{Limitations}
\label{sec:limitations}
\paragraph{Gold-standard scope.}
All reported F1 and prevalence-gap metrics use a stratified gold-standard set of 1,698 items (2-of-3 majority reference labels), not the full 50{,}087-text corpus. This follows the expert-reference evaluation design in OATH-Frames \citep{ranjit2024oath} and the practice of using manually annotated gold standards in low-resource NLP \citep{cardoso2014gold}: the gold standard is a \emph{reference} for auditing models, not a claim of infallible truth on subjective bias. Rare categories (e.g., \textit{racist}, 5 gold positives under 2-of-3) have too few positives for stable slice-level prevalence gaps (Appendix~\ref{sec:racist_exclusion}, Table~\ref{tab:prevalence_gap_ci}). Corpus-wide GPT-4.1 labels are exploratory and can drift substantially from reference prevalence by source (Table~\ref{tab:label_prevalence_gold_vs_gpt}).

\paragraph{Coverage and geography.}
Drawing text from Reddit, X, news, and city council minutes does not encompass all homelessness discourse. We do not verify whether posts are human- or bot-authored. Collection is confined to ten U.S.\ cities and English text chosen for stratified comparison; they do not represent all socio-economic or cultural contexts nationally or globally. Our 11-term PEH lexicon is shared across cities and may miss locally specific vocabulary.

\paragraph{Subjectivity and deployment.}
Despite partner-guided taxonomy design and three-annotator soft labels \citep{fornaciari2021beyond}, implicit stigma without overt slurs remains hard for humans and LLMs to code consistently. Moderate macro-F1 can coexist with large prevalence gaps (Section~\ref{sec:ro3_calibration}); municipalities should not treat LLM corpus dashboards as ground truth without recurring gold-standard audits (operational guidance in Appendix~\ref{sec:municipal_deployment}). Vote-threshold retuning can reduce NIMBY gaps (Appendix~\ref{sec:nimby_mitigation}); pseudo-label fine-tuning can compress gaps further on gold (Appendix~\ref{sec:evaluation_framing}) but inherits teacher error patterns on other labels and does not make fully automated enforcement safe.

\paragraph{Annotation and city coverage.}
Annotator training, backgrounds, and sparse-cell $\kappa$ interpretation are detailed in Appendix~\ref{sec:annotator_agreement}. Disagreement is highest for \textit{provide fact/claim} on council speech (Table~\ref{tab:disagreement_fact_claim}). Kalamazoo and Baltimore lack usable verbatim council transcripts; council-modality findings exclude those cities (Appendix~\ref{sec:data}).

\paragraph{NIMBY error patterns.}
Section~\ref{sec:nimby_error_main} and Appendix~\ref{sec:nimby_error_analysis} report co-occurring lexical cues on consensus false positives; Appendix~\ref{sec:nimby_representation} adds embedding probes and counterfactual edits. These analyses are descriptive, not causal LLM ablations. On the gold-standard set, local Qwen NIMBY prevalence gaps remain positive under prompt variants A/B/C ($+37.5$, $+10.2$, $+39.8$\,pp; Table~\ref{tab:nimby_prompt_sensitivity}).

\section*{Ethical Statement}
The principle of beneficence, which maximizes benefits while minimizing potential harms \citep{beauchamp2008principle}, is critical to our research. Promoting fairness is especially important when studying biases against PEH, a socially vulnerable population. We have worked in close collaboration with specialized non-profits in South Bend to guide the manual annotation of homelessness biases.

Privacy is a central concern. All data are anonymized to remove PII using spaCy and pydeidentify, adhering to ethical standards for data privacy and ensuring that individuals’ identities are protected. As an additional check, we spot-checked all of the gold standard, ensuring that there was no PII.

In creating the paper, we used LLMs such as ChatGPT and Gemini to help with coding and writing, though all of the content is our own.

\section*{Acknowledgments}
This project is a collaborative effort involving the City of South Bend, the University of Notre Dame Center for Social Concerns, and local non-profits. We would like to thank Dr. Margaret Pfeil, co-founder of Motels4Now, a local non-profit that provides housing for PEH, for her guidance in this project. Additionally, Marshall Smith, who works at Broadway Christian Parish and the City of South Bend, provided great support throughout this process. We also thank the University of Notre Dame for the Strategic Framework Grant that makes this work possible. Finally, we thank Emory Smith and Lina McKimson, who have worked with the Lucy Family Institute for Data \& Society.

\renewcommand\refname{References}
\bibliography{custom}

\appendix
\renewcommand{\thesection}{\AlphAlph{\value{section}}}
\section{Related Work}

\subsection{LLMs for Text Classification and Policy Research}
Recent work demonstrates that LLMs perform well on low-resource classification tasks through zero-shot and few-shot learning \citep{matarazzo2025survey}, and are increasingly used in social science to code large datasets and extract policy-relevant information \citep{gilardi2023chatgpt, halterman2024codebook}. However, LLMs themselves contain biases that can influence classification outcomes \citep{li2025understanding, blodgett2020language}, requiring careful evaluation when used for bias detection tasks. A recent benchmark study \citep{majumdar2025evaluating} evaluates LLMs as demographic-targeted bias detectors across nine axes, twelve English datasets, and zero-, few-shot, and fine-tuned settings, finding strong fine-tuned smaller models but persistent gaps on multi-target and intersectional cases.

While we focus on LLMs to detect human-generated bias, it is crucial to acknowledge the inherent biases within LLMs themselves (e.g., representational biases, harmful content generation). Techniques for auditing LLM outputs for fairness between demographic groups or identifying stereotypical associations within their internal representations \citep{bolukbasi2016man, nadeem2020stereoset} are relevant to ensure the integrity of classification results. This consideration motivates our use of human-annotated gold standards for evaluation rather than relying solely on LLM agreement.

\subsection{LLM Annotations, Prevalence, and Human Calibration}
Computational social science increasingly uses LLMs to code text at scale \citep{gilardi2023chatgpt, halterman2024codebook}, but agreement metrics such as F1 or Cohen's $\kappa$ can hide \emph{label prevalence shift}: a model may partially overlap human positives while systematically over or under-predicting a category \citep{fornaciari2021beyond,card2020little}. Prior work on hate-speech annotation further shows that subjective labels vary with annotator target socio-demographic alignment and that persona-based LLM annotators can exhibit bias patterns that differ from humans \citep{giorgi2025human}.

For subjective categories, soft labels rather than a single hard label when annotators disagree are standard in subjective toxicity coding \citep{fornaciari2021beyond}, with extensive discussion of when majority vote is appropriate for multi-rater reference labels \citep{artstein2008survey,davani2022dealing}. We adopt the same spirit on our gold-standard set: three coders per item, 2-of-3 majority for binary reference labels, with sensitivity to a lenient any-vote cut reported in the calibration appendix.

Overlap metrics alone do not answer whether a monitor changes estimated prevalence. We therefore report \textbf{prevalence gaps} following work on soft labels and small-refinement effects in subjective annotation \citep{fornaciari2021beyond,card2020little}, because municipal partners care whether a monitor \emph{inflates} NIMBY rates, not only whether it attains moderate F1. When multiple LLMs are ensembled, aggregation rules such as vote-fraction thresholds are part of the monitor specification, not neutral defaults \citep{zheng2023judging}.

Corpus-scale teacher labels inherit reference drift: PEH framing work separates expert reference annotations from scaled LLM predictions \citep{ranjit2024oath}, and LLM coders can diverge from human prevalence even at moderate agreement \citep{gilardi2023chatgpt}. We treat GPT-4.1 corpus labels as exploratory unless checked against the gold-standard set on a standing audit sample.

\subsection{AI for Detecting Societal Bias}
\label{sec:ai_societal_bias}

Previous studies have applied LLMs to detect biases against the poor (aporophobia) in online discourse \citep{kiritchenko2023aporophobia, curto2024crime, curto2025tackling}. For PEH specifically, OATH-Frames \citep{ranjit2024oath} releases expert-only reference annotations (4.1K posts), LLM-assisted expert annotations, and model-predicted labels at scale. It distinguishes evaluation references from scaled teacher labels, as we do for gold-standard vs.\ GPT-4.1 corpus labels. OATH categorizes biases into nine frames including `not in my backyard', `harmful generalization', and `deserving/undeserving'; LLM-classified tweets correlate unsheltered PEH population size with harmful generalizations \citep{ranjit2024oath}.

However, OATH was limited to a single platform (X) and a single keyword (`homeless'). Our research advances this area by: (1) collecting multi-domain data from Reddit, X, news, and text transcribed from city council meetings; (2) utilizing a comprehensive PEH lexicon with 11 terms \citep{karr2025homelessness} (see Appendix \ref{sec:peh_lexicon}); and (3) expanding the taxonomy to 16 categories to capture nuanced biases across platforms, as detailed in Section \ref{sec:bias_classification}.

\section{PEH Lexicon}
\label{sec:peh_lexicon}
The corpus uses the following 11 search terms \citep{karr2025homelessness}: `homeless', `homelessness', `housing crisis', `affordable housing', `unhoused', `houseless', `housing insecurity', `beggar', `squatter', `panhandler', and `soup kitchen'.

\section{Data Anonymization Details}
\label{sec:anonymization_details}
The Methodology section describes anonymization before annotation and model inference; this appendix documents the entity types and tools used.
We leveraged the spaCy NLP library \citep{HonnibalMontani2020} to automatically identify and mask personally identifiable information (PII) within the text. The specific categories of entities targeted for anonymization included: person names, geographic locations, organizations, and other identifying information such as street addresses, phone numbers, and emails. We also used the Python module pydeidentify \citep{pydeidentify}, which is based on spaCy, as a secondary pass to ensure comprehensive anonymization.

\section{Data}
\label{sec:data}
This appendix supplements the city-selection and corpus-scale discussion in the Corpus section (Table~\ref{tab:summary_data} in the main text). It details the kNN matching procedure, county covariates, and per-city collection counts.

\subsection{Collection and labeling pipeline}
\label{sec:collection_pipeline}
\begin{figure}[t]
    \centering
    \includegraphics[width=0.85\linewidth]{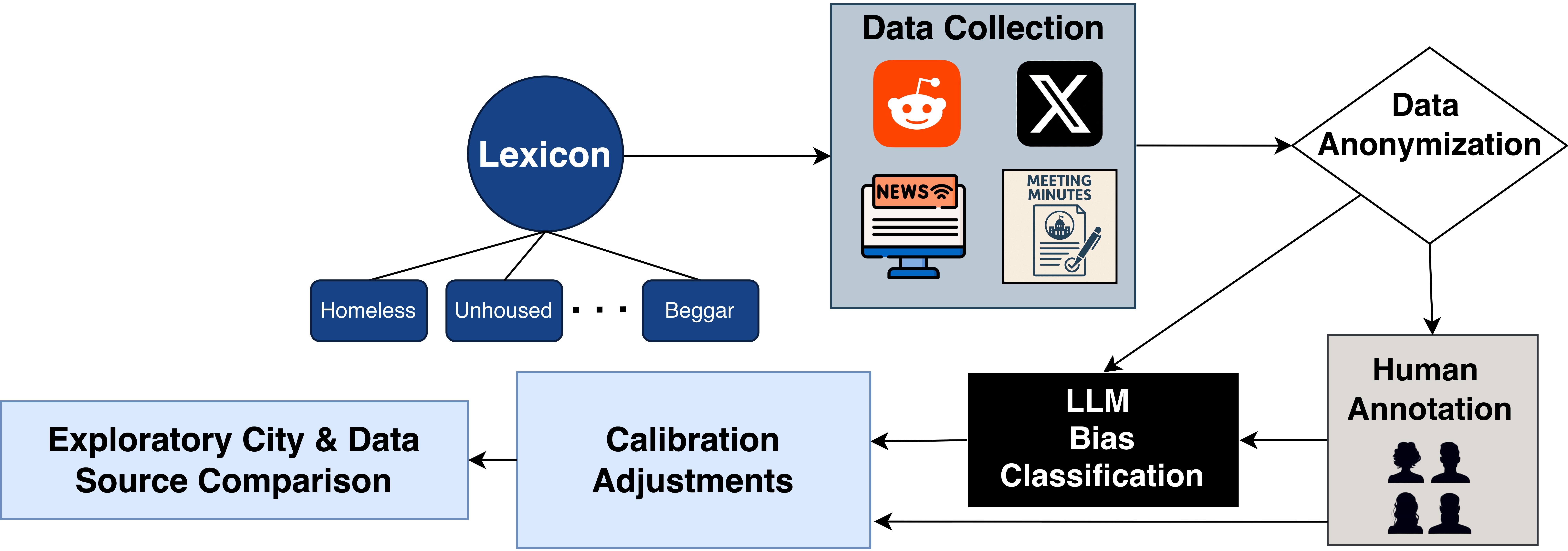}
    \caption{Multi-domain PEH bias corpus pipeline: lexicon retrieval, anonymization, stratified gold-standard annotation (human reference), GPT-4.1 teacher labeling of the remaining corpus (\emph{not} ground truth), and prevalence-gap audits against gold before any exploratory corpus-scale reporting.}
    \label{fig:pipeline}
\end{figure}

Figure~\ref{fig:pipeline} shows the end-to-end workflow referenced in Section~\ref{sec:corpus}: lexicon-based retrieval from four sources, spaCy anonymization, stratified gold-standard annotation (1,698 items), GPT-4.1 few-shot labeling of the remaining corpus, and prevalence-gap audits against gold reference labels before any corpus-scale reporting.

We selected ten U.S.\ cities, grouped into two sets of five, based on our kNN script. When selecting the cities, we gathered data from their respective counties from the U.S.\ Census. The counties for these cities were selected to have similar levels of population, homelessness rates, and Gini index, while differing primarily in racial fragmentation. The cities of South Bend, Indiana, and San Francisco, California, were selected as the cities for each group.

Our dataset from the ten cities can be seen in Table~\ref{tab:summary_data}. The dataset comprises 50,087 total text samples collected between January 1, 2015, and January 1, 2025, with the largest contributions from Reddit comments (34,486) and X non-reposts (3,843). Smaller cities like Scranton and El Paso contribute fewer total data samples than larger cities like San Francisco and Portland. Kalamazoo and Baltimore do not publish usable verbatim council speech (only agendas or short recaps), so council-comment counts are denoted N/A for those cities. Additionally, certain cities have a lower number of meetings since they do not span ten years. For example, the city of San Francisco began publishing full meeting transcripts in 2023; prior to that, only summaries were available. The given distribution reflects both population differences and varying levels of online civic engagement across cities.

\section{X (Twitter) User Geolocation Inference}
\label{sec:geolocation}
The corpus section describes our geolocation pipeline; this appendix reports coverage and geographic discourse patterns for the 2025 X subsample.

\subsection{X Geolocation Results}
\label{sec:x_geolocation_main}
Our 2025 X subsample is geolocated with the multi-stage pipeline below. Roughly 80\% of users receive a location; external ``star'' accounts ($>$1000 followers) account for most out-of-state reach, while Baltimore shows more local regular users than West Coast cities (Tables~\ref{tab:geolocation_results} and~\ref{tab:geographic_patterns}).

\noindent Fewer than 1\% of posts contain GPS coordinates. We infer user location $l_i$ by a multi-stage fallback:
\begin{equation}
l_i = \begin{cases}
    f_{\text{self}}(u_i) & \text{if valid self-report} \\
    f_{\text{post}}(u_i) & \text{else if post-derived signal} \\
    f_{\text{friend}}(u_i) & \text{else if friend-based inference} \\
    \emptyset & \text{otherwise}
\end{cases}
\end{equation}
\noindent Stages: (1) \textbf{Self-reported} profile geocoding (city-level or finer); (2) \textbf{Post-derived} geo-tagged places or text mentions; (3) \textbf{Star-user} handling (followers $>1000$, national audience); (4) \textbf{Friend-based} homophily for regular users. Each stage assigns high/medium/low confidence.

\subsection{Coverage and discourse patterns}
\label{sec:geolocation_appendix_results}

Due to API cost, the X subset spans 2025 only. Table~\ref{tab:geolocation_results} summarizes geolocation inference by city (self-reported profile text, post-derived signals, friend-network inference).

\begin{table}[htpb!]
\centering
\resizebox{\columnwidth}{!}{
\begin{tabular}{lrrrrr}
\toprule
\textbf{City} & \textbf{Total} & \textbf{Self-Rep.} & \textbf{Post-Der.} & \textbf{Friend} & \textbf{Bot/None} \\
\midrule
Portland & 3,880 & 1,331 & 551 & 1,247 & 751 \\
San Francisco & 5,112 & 1,856 & 723 & 1,689 & 844 \\
Baltimore & 1,245 & 489 & 187 & 412 & 157 \\
Buffalo & 1,089 & 398 & 156 & 378 & 157 \\
South Bend & 281 & 112 & 21 & 98 & 50 \\
\midrule
\textbf{Coverage} & - & 34.2\% & 14.2\% & 32.1\% & 19.5\% \\
\bottomrule
\end{tabular}
}
\caption{Geolocation inference by city. Self-Rep.\ = self-reported; Post-Der.\ = post-derived; Friend = friend-based inference.}
\label{tab:geolocation_results}
\end{table}

Across cities, the pipeline geolocates $\approx$80.5\% of users: 34.2\% self-reported, 14.2\% post-derived, 32.1\% friend-based; 19.5\% remain unlocated (flagged as potential bots). Table~\ref{tab:geographic_patterns} summarizes discourse patterns.

\begin{table}[htpb!]
\centering
\resizebox{\columnwidth}{!}{
\begin{tabular}{lrrr}
\toprule
\textbf{Pattern} & \textbf{Portland} & \textbf{San Francisco} & \textbf{Baltimore} \\
\midrule
Local regular users & 20.7\% & 14.7\% & 33.8\% \\
External star influence & 94.2\% & 84.7\% & 86.7\% \\
Texas star users & 82 & 99 & 1 \\
NY star users & 101 & 107 & 1 \\
\bottomrule
\end{tabular}
}
\caption{Geographic patterns. Local regular users = \% of regular users in target city; external star influence = \% of star-user followers from out-of-state.}
\label{tab:geographic_patterns}
\end{table}

\begin{itemize}
    \item \textbf{External amplification}: 85-95\% of star-user influence is out-of-state; locals generate most posts but external accounts dominate reach.
    \item \textbf{Texas vs.\ Baltimore}: Texas-based stars focus on West Coast cities (82-99 users) vs.\ one for Baltimore; NY stars are nationwide ($\sim$100 users per major city).
    \item \textbf{Baltimore is more local} (33.8\% local regular users vs.\ 14.7-20.7\% on the West Coast); smaller cities have almost no star users.
\end{itemize}

\subsection{Multi-Stage Inference}

\subsubsection{Stage 1: Self-Reported Location Extraction}

The first stage extracts location information from users' profile metadata.
X allows users to specify a free-text location field in their profile, which
may contain precise locations (e.g., ``San Francisco, CA''), ambiguous
descriptions (e.g., ``Bay Area''), or non-geographic text (e.g., ``Planet Earth'').

\paragraph{Geocoding.} We employ the Google Maps Geocoding API to parse
free-text location strings into structured geographic coordinates and
normalized place names. Each location string $s$ is mapped to a tuple
$(p, \text{level})$ where $p$ is the parsed location name and $\text{level}
\in \{\texttt{city}, \texttt{state}, \texttt{country}, \texttt{invalid}\}$
indicates the geographic granularity.

\paragraph{Validation.} To ensure high confidence, we retain only city-level
or finer locations. Specifically, we filter out:
\begin{itemize}
    \item \textbf{Country-level locations}: ``United States'', ``USA'', ``Canada'', etc.
    \item \textbf{State-only locations}: ``California'', ``CA'', ``Texas'', etc. (without city specification)
    \item \textbf{Non-geographic strings}: ``Earth'', ``Worldwide'', ``Internet'', etc.
\end{itemize}

\noindent Users with valid city-level locations are assigned $c_i = \texttt{high}$
and excluded from subsequent processing stages.

\subsubsection{Stage 2: Post-Based Location Inference}

For users without valid self-reported locations, we extract location signals
from their posting behavior. We identify three signal types with decreasing reliability:

\paragraph{(a) Geo-tagged Places.} When users enable location sharing, X
attaches a \texttt{place} object to their posts containing the location name.
This represents the most reliable post-derived signal as it reflects the
user's actual posting location. For users with multiple geo-tagged posts,
we select the most frequent location:

\begin{equation}
l_i^{\text{geo}} = \argmax_{p \in P_i} |\{t \in T_i : \text{place}(t) = p\}|
\end{equation}

\noindent where $T_i$ is the set of user $i$'s posts and $P_i$ is the set of distinct
places. Users located via geo-tags receive $c_i = \texttt{high}$.

\paragraph{(b) Network Locations.} We extract locations from users' social
interactions, specifically from mentioned users' profile locations and
replied-to users' profile locations. The intuition is that users frequently
interact with geographically proximate individuals. To reduce noise, we
require a location to appear in at least two distinct interactions:

\begin{equation}
l_i^{\text{net}} = \argmax_{p \in P_i^{\text{net}}} \text{count}(p)
\quad \text{s.t.} \quad \text{count}(p) \geq 2
\end{equation}

\noindent Users located via network signals receive $c_i = \texttt{medium}$.

\paragraph{(c) Text-Mentioned Places.} We leverage AI-detected place mentions
in post text, extracted using named entity recognition. Given the high false
positive rate (users may discuss locations they do not reside in), we apply
strict filtering: probability threshold $> 0.9$ and occurrence count $\geq 2$.
Users located via text mentions receive $c_i = \texttt{low}$.

\subsubsection{Stage 3: Star vs. Regular User Classification}

We classify users into two categories based on follower and following counts.
Star users (news media, public figures, and organizations) have nationwide or global audiences, making friend-based geolocation less reliable. They also
require costly API calls to retrieve large follower lists:

\begin{equation}
\begin{aligned}
\text{type}(u_i) =
\begin{cases}
\texttt{star}, &
\begin{aligned}
&\text{if } |\text{followers}_i| > \tau \\
&\land\ |\text{following}_i| > \tau
\end{aligned}
\\
\texttt{regular}, & \text{otherwise}
\end{cases}
\end{aligned}
\end{equation}

\noindent where $\tau = 1000$ is the classification threshold. 
We apply differentiated processing strategies as shown in
Table~\ref{tab:user_classification}.

\begin{table}[t]
\centering
\small
\begin{tabular}{lll}
\toprule
\textbf{User Type} & \textbf{Condition} & \textbf{Processing Strategy} \\
\midrule
Regular & default & Full friend list retrieval \\
Star & followers $\leq$ 5,000 & Limited to first 200 friends \\
Star & followers $>$ 5,000 & Excluded (cost prohibitive) \\
\bottomrule
\end{tabular}
\caption{Differentiated processing strategies for user types based on follower count thresholds.}
\label{tab:user_classification}
\end{table}

\subsubsection{Stage 4: Social Network-Based Geolocation}

For users not located in previous stages, we infer location from the
geographic distribution of their social connections. 
This approach relies on geographic homophily: users tend to follow and be
followed by others who live nearby.

\paragraph{Friend Location Collection.} For each user $u_i$, we retrieve
their follower set $F_i$ and following set $G_i$ via the X API. For each
friend $f \in F_i \cup G_i$, we extract and geocode their profile location
using the same process as Stage 1.

\paragraph{Location Estimation.} We assign the most frequent location among
friends:

\begin{equation}
l_i^{\text{friend}} = \argmax_{p \in L(F_i \cup G_i)} |\{f \in F_i \cup G_i : l_f = p\}|
\end{equation}

\noindent where $L(\cdot)$ denotes the set of geocoded locations.

\paragraph{Confidence Assignment.} Based on the geographic granularity of
the inferred location, we assign confidence levels as follows:

\begin{equation}
c_i = \begin{cases}
    \texttt{medium} & \text{if } \text{level}(l_i^{\text{friend}}) = \texttt{city} \\
    \texttt{low} & \text{if } \text{level}(l_i^{\text{friend}}) \in \{\texttt{state}, \texttt{country}\} \\
    \texttt{none} & \text{if } l_i^{\text{friend}} = \emptyset \text{ (likely bot)}
\end{cases}
\end{equation}

\paragraph{Bot Detection.} Users with no locatable friends or highly
dispersed friend locations (no dominant location) are flagged as potential
bot accounts and assigned $c_i = \texttt{none}$.

\subsubsection{Stage 5: Priority-Based Result Merging}

The final stage merges results from all sources using a strict priority
ordering that reflects signal reliability:

\begin{equation}
\text{Priority}: \quad \underbrace{f_{\text{self}}}_{\text{highest}} >
\underbrace{f_{\text{geo}}}_{\text{high}} >
\underbrace{f_{\text{net}}}_{\text{medium}} >
\underbrace{f_{\text{text}}}_{\text{low}} >
\underbrace{f_{\text{friend}}}_{\text{lowest}}
\end{equation}

\noindent For each user, we select the location from the highest-priority source
that yields a non-empty result. This ensures that higher-confidence
estimates are never overwritten by lower-confidence ones.

\subsection{Implementation Details}

\paragraph{API Cost Management.} The Google Maps Geocoding API charges
\$5 per 1,000 requests. To minimize costs, we implement an aggressive
caching strategy that stores all geocoding results. The cache is shared
across cities and processing runs, reducing redundant API calls by
approximately 85\%.

\paragraph{Incremental Processing.} The pipeline supports checkpoint-based
resumption. Results are saved every 10 processed users, enabling recovery
from interruptions without data loss.

\paragraph{Parallel Processing.} For large cities (e.g., Portland with
1,757 users, San Francisco with 2,544 users), Stage 4 supports chunked
processing via the \texttt{-chunk N/M} parameter, enabling distributed
execution across multiple machines.

\section{Classification Taxonomy}
\label{sec:classification_categories}
We expanded upon the nine OATH-Frames \citep{ranjit2024oath} to create 16 classification categories (Section~\ref{sec:bias_classification}); Table~\ref{tab:taxonomy} below gives full definitions for reproducibility. The additional categories allow us to address new dataset challenges, such as distinguishing between genuine and rhetorical questions commonly found on Reddit. We created the \textit{Negative Bias Frame}, which has the five categories `ask a rhetorical question', `not in my backyard', `harmful generalization', `deserving/undeserving', and `racist' (Appendix~\ref{sec:exploratory_corpus}).

\begin{table*}[htpb!] 
\centering
\small
\begin{tabularx}{\textwidth}{@{} l l X @{}}
\toprule
\textbf{Major Category} & \textbf{Sub-Category} & \textbf{Description} \\ \midrule
\rowcolor[gray]{0.95} \multicolumn{3}{l}{\textit{OATH Categories}} \\
\multirow{3}{*}{Critique} & Money Aid Allocation & Discussion of financial resources, aid distribution, or resource allocation. \\
 & Government Critique & Criticism of government policies, laws, or political approaches. \\
 & Societal Critique & Criticism of social norms, systems, or societal attitudes. \\ \addlinespace
Response & Solutions/Interventions & Discussion of specific solutions, interventions, or charitable actions. \\ \addlinespace
\multirow{5}{*}{Perception} & Personal Interaction & Direct personal experiences with persons experiencing homelessness (PEH). \\
 & Media Portrayal & Discussion of PEH as portrayed in media. \\
 & NIMBY & Opposition to local homelessness developments (Not In My Backyard). \\
 & Harmful Gen. & Negative stereotypes or sweeping statements about PEH. \\
 & Deserving/Undeserving & Judgments regarding who deserves assistance. \\ \midrule
\rowcolor[gray]{0.95} \multicolumn{3}{l}{\textit{New Categories}} \\
\multirow{6}{*}{Comment Types} & Genuine Question & The speaker asks a sincere question about homelessness issues. \\
 & Rhetorical Question & A question not intended to be answered, often used to make a point. \\
 & Fact or Claim & Provides a factual statement or claim about homelessness. \\
 & Observation & Shares an observation about homelessness or related situations. \\
 & Express Own Opinion & The speaker expresses their own views or feelings. \\
 & Express Others' Op. & Describes or references the views or feelings of others. \\ \addlinespace
Racist & Racist & Contains explicit or implicit racial bias. \\ \bottomrule
\end{tabularx}
\caption{Classification Taxonomy}
\label{tab:taxonomy}
\end{table*}

\begin{table}[t]
\centering
\scriptsize
\setlength{\tabcolsep}{2pt}
\begin{tabularx}{\columnwidth}{@{}>{\raggedright\arraybackslash}X r r r r@{}}
\toprule
\textbf{Label} & \multicolumn{2}{c}{\textbf{Gold}} & \multicolumn{2}{c}{\textbf{GPT}} \\
 & \% & $n_{+}$ & \% & $n_{+}$ \\
\midrule
Genuine Q & 6.7 & 113 & 11.3 & 5670 \\
Rhetorical Q & 8.7 & 148 & 9.3 & 4656 \\
Fact/claim & 83.4 & 1416 & 74.3 & 37281 \\
Observation & 8.7 & 148 & 35.1 & 17601 \\
Own opinion & 53.4 & 907 & 83.1 & 41695 \\
Others' op. & 6.5 & 111 & 17.9 & 8972 \\
Money/aid & 18.2 & 309 & 23.1 & 11586 \\
Gov. critique & 12.0 & 203 & 29.3 & 14694 \\
Soc. critique & 9.0 & 152 & 62.7 & 31440 \\
Solutions & 39.9 & 678 & 34.6 & 17347 \\
Personal & 4.7 & 80 & 13.5 & 6776 \\
Media & 0.6 & 10 & 6.9 & 3474 \\
NIMBY & 2.5 & 42 & 8.6 & 4290 \\
Harm. gen. & 6.9 & 118 & 24.5 & 12303 \\
Deservingness & 1.4 & 23 & 27.6 & 13849 \\
Racist & 0.3 & 5 & 0.1 & 69 \\
\bottomrule
\end{tabularx}
\caption{Label prevalence (\%) and $n_{+}$ for 16 categories. Gold: $n{=}1{,}698$ ($\tau{=}2/3$); GPT corpus: $n{=}50{,}161$ (exploratory).}
\label{tab:label_prevalence_gold_vs_gpt}
\end{table}

\section{Why These Categories and the Negative Bias Frame}
\label{sec:taxonomy_rationale}
Section~\ref{sec:bias_classification} introduces the 16-label schema and Negative Bias Frame; here we justify both design choices for policy partners.

\paragraph{Why a 16-category multi-label schema.}
Homelessness discourse is not a single ``pro vs.\ anti'' stance. Partners in South Bend needed to distinguish \emph{how} bias appears: inflammatory rhetoric (\textit{ask a rhetorical question}), local opposition to services (\textit{not in my backyard}), stereotyping (\textit{harmful generalization}), moral deservingness judgments, explicit racism, but also constructive strands (\textit{solutions/interventions}, \textit{provide fact/claim}). Collapsing to sentiment or a single toxicity score would hide whether a community opposes shelters, blames PEH, or debates policy on evidence. The OATH frames \citep{ranjit2024oath} provided a core for X-centric stigma labels; we extended them for Reddit questions, factual vs.\ observational claims, and attributed opinions because those forms dominated council and social text in pilot annotation.

\paragraph{Why the Negative Bias Frame uses these five categories.}
The frame is intentionally narrow (not all 16 labels), so this monitoring targets \emph{actionable stigma} rather than all critical speech:
\begin{itemize}[nosep]
    \item \textbf{Ask a rhetorical question:} Partners flagged sarcastic questions as a common device that delegitimizes PEH without explicit slurs; it is prevalent online and separable from genuine information-seeking.
    \item \textbf{Not in my backyard:} Directly motivated our South Bend collaboration (shelter siting backlash); highest engagement in our Reddit analysis (Appendix~\ref{sec:reddit_engagement_appendix}).
    \item \textbf{Harmful generalization} and \textbf{deserving/undeserving:} Core dehumanizing and blame frames in prior PEH bias work \citep{ranjit2024oath,curto2025tackling}; strongly concentrated on social media in our corpus.
    \item \textbf{Racist:} Recommended by local stakeholders for intersectional bias against PEH; retained in the frame for prevalence and engagement analyses despite extreme sparsity (see Appendix~\ref{sec:racist_exclusion} for prevalence-gap exclusion).
\end{itemize}
We \emph{exclude} labels such as \textit{societal critique}, \textit{government critique}, and \textit{express opinion} from the frame because they frequently co-occur with neutral reporting or constructive advocacy; including them would inflate ``bias'' rates and blur monitoring of stigma aimed at PEH populations.

\section{Racist Label: Sparsity and Calibration Gap Exclusion}
\label{sec:racist_exclusion}
Section~\ref{sec:ro3_calibration} omits \textit{racist} from prevalence-gap tables but retains it elsewhere; this appendix explains why.

\paragraph{Rare but important.}
\textit{Racist} appears in $<$1\% of corpus-wide instances and has the highest annotator unanimity ($\sim$97\%, Appendix~\ref{sec:annotator_agreement}) when it does occur; annotators agree it is rare, not that it is ambiguous. It remains in the full taxonomy, F1 benchmarks, and Negative Bias Frame prevalence/engagement analyses because explicit racial bias in homelessness discourse is policy-relevant even when infrequent.

\paragraph{What counts as racist here (stakeholder definition).}
Partners asked us to separate \textit{racist} from generic stigma. We code \textbf{Yes} when the text invokes race/ethnicity to blame, exclude, or demean PEH or to contrast racialized groups in resource allocation (e.g., ``\ldots refuse to deport migrants + give them Black taxpayers' [money] 4 shelter+food while Black citizens go homeless''). Explicit slurs are positive but uncommon. \textbf{Implicit / coded} cases include race-coded tropes that link homelessness to a racialized out-group without naming a slur (e.g., blaming ``illegals'' for housing scarcity while attacking PEH services, or pitting racialized constituencies against each other around shelters). We distinguish this from (i) \textit{prejudice} expressed as non-racial stereotyping of PEH (often \textit{harmful generalization} / \textit{deserving/undeserving}) and (ii) \textit{discrimination} as reported policy outcomes without racial framing. Positive examples are sparse on gold (5 under 2-of-3), so prevalence-gap tables omit the label while F1 still includes it where support allows.

\paragraph{Why exclude from prevalence-gap tables.}
Prevalence gaps $\hat{\pi}_{\mathrm{model}}-\pi_{\mathrm{ref}}$ require stable estimates of reference positive rate $\pi_{\mathrm{ref}}$ per slice. On the gold-standard set ($n{=}1{,}698$), there are \textbf{only five} positive soft-label instances for \textit{racist} (2-of-3 majority). With so few positives, a handful of model errors shifts gaps by tens of percentage points, producing misleading ``calibration'' numbers. We therefore exclude \textit{racist} from prevalence-gap summaries while still reporting model behavior on the label in standard F1 evaluation where support allows. This is a statistical support decision, not a claim that racism is unimportant.

\FloatBarrier
\section{Inter-Annotator Agreement}
\label{sec:annotator_agreement}
Section~\ref{sec:bias_classification} reports mean annotator unanimity (78.38\%) and soft-label construction; this appendix provides per-category agreement rates and Cohen's $\kappa$ against majority-vote gold.

To establish the reliability of our multi-label classification, we conducted an Inter-Annotator Agreement (IAA) study on the 1,698-item gold-standard set with three annotators trained by partners from Motels4Now and Broadway Christian Parish using a coding guide aligned with the 16-category taxonomy (definitions in Appendix~\ref{sec:classification_categories}). Training included guide review, pilot coding on $\approx$20 items per source, and adjudication sessions with domain experts before independent labeling. Annotators had mixed backgrounds (graduate researchers and partner staff with homelessness-policy experience); disagreements were \textbf{not} resolved by discussion on the audit set; reference labels use 2-of-3 majority vote (soft scores $s_{ij}\in\{0,\frac{1}{3},\frac{2}{3},1\}$) for evaluation \citep{fornaciari2021beyond}.

The gold standard was created via stratified sampling across ten cities and four sources (up to 50 entries per city $\times$ source; Appendix~\ref{sec:data}). Table~\ref{tab:humans_vs_gold_kappa} reports Cohen's $\kappa$ between a random annotator and majority-vote gold by source $\times$ category. Asterisks mark cells with \textbf{$<5$ gold positives}: $\kappa{=}0.00$ there reflects \emph{empty or near-empty prevalence}, not chance agreement on abundant labels. For example, news has \textbf{zero} gold NIMBY positives in the audit set ($n{=}382$ news items), so news-NIMBY $\kappa$ is undefined/degenerate; racist has 0 news and 0 meeting positives. Reddit and X carry most NIMBY and stigma labels; council and news cells are more reliable for factual and policy-framing categories. Mean per-cell unanimity is 78.38\%.

\paragraph{Disagreement examples: fact/claim.}
Council transcripts and news blur fact vs.\ opinion (lowest meeting-minutes fact/claim $\kappa{=}0.31$). Table~\ref{tab:disagreement_fact_claim} lists de-identified excerpts with soft scores $1/3$ or $2/3$, illustrating where humans split before any LLM is scored.
Table~\ref{tab:iaa_soft_structure} and Table~\ref{tab:iaa_kappa_mechanisms} show that lower $\kappa$ tracks split votes and/or rarity, not uniform noise.

\begin{table}[t]
\centering
\footnotesize
\setlength{\tabcolsep}{3pt}
\begin{tabularx}{\columnwidth}{@{}l c >{\raggedright\arraybackslash}X@{}}
\toprule
\textbf{Source} & $s_{ij}$ & \textbf{De-identified excerpt} \\
\midrule
council & $1/3$ & ``And we left one term, excuse me, one position vacant on that, that was also ending [ORGANIZATION]. [ORGANIZATION], [ORGANIZATION], ORG2, one term involving housing services. [ORGANIZATION], one term involving homelessness services.'' \\
council & $1/3$ & ``I just have a couple of questions. You mentioned, is this just to blanket homeless or you said the general public? The general public representative.'' \\
council & $2/3$ & ``But I think that a bathhouse would be a relatively productive thing to put in certain parts of the city. I don't know how that would be done. But it does give -- makes it a little more human for the homeless.'' \\
council & $2/3$ & ``However, we feel that the \$500 that was discussed would not be an issue as far as eligibility if the [ORGANIZATION] was to be dedicated or allocated towards this particular program and So again, we did look at a couple of different sources. We looked at the\ldots'' \\
news & $2/3$ & ``Dr PERSON0. For services to [PERSON] and to the community in [ORGANIZATION], [PERSON], particularly during [ORGANIZATION] [ORGANIZATION], [PERSON]'' \\
reddit & $1/3$ & ``There are a lot of people who want to be nice to the homeless as a concept because they don't have to deal with it. As an example, I have a relative who is like that. Wants `something nice done' for the homeless. I asked her if she wanted to deal with a\ldots'' \\
\bottomrule
\end{tabularx}
\caption{Illustrative annotator disagreements on \textit{provide a fact or claim} (soft score $s_{ij}\in\{1/3,2/3\}$), emphasizing council minutes where procedural, evaluative, and asserted-claim speech co-occur. Raters often split on whether an utterance is a verifiable claim vs.\ opinion, clarification question, or agenda meta-discourse (the same ambiguity that drives low fact/claim $\kappa$ on this source).}
\label{tab:disagreement_fact_claim}
\end{table}

\begin{table}[!htb]
\centering
\scriptsize
\setlength{\tabcolsep}{2pt}
\begin{tabularx}{\columnwidth}{@{}>{\raggedright\arraybackslash}X r r r r@{}}
\toprule
\textbf{Cat.} & $n_{+}$ & Split & Unan. & Prev. \\
\midrule
Solutions & 678 & 57.2 & 42.8 & 39.8 \\
Own opinion & 907 & 48.3 & 51.7 & 53.3 \\
Money/aid & 309 & 32.3 & 67.6 & 18.2 \\
Fact/claim & 1420 & 29.9 & 70.0 & 83.4 \\
Soc. critique & 152 & 29.0 & 71.0 & 8.9 \\
Others' op. & 112 & 25.9 & 74.1 & 6.6 \\
Gov. critique & 203 & 23.7 & 76.3 & 11.9 \\
Observation & 148 & 22.1 & 77.9 & 8.7 \\
Harm. gen. & 118 & 19.5 & 80.4 & 6.9 \\
Personal & 80 & 10.5 & 89.5 & 4.7 \\
NIMBY & 42 & 10.2 & 89.8 & 2.5 \\
Rhetorical Q & 148 & 10.0 & 90.0 & 8.7 \\
Genuine Q & 113 & 9.2 & 90.8 & 6.6 \\
Deservingness & 23 & 8.4 & 91.6 & 1.4 \\
Media & 11 & 6.9 & 93.1 & 0.6 \\
Racist & 5 & 2.6 & 97.4 & 0.3 \\
\bottomrule
\end{tabularx}
\caption{Soft-label structure ($n{=}1{,}698$; 2-of-3 gold). Split = share with $s_{ij}\in\{1/3,2/3\}$; Unan.\ = $s_{ij}\in\{0,1\}$; Prev.\ = 2-of-3 positive rate.}
\label{tab:iaa_soft_structure}
\end{table}

\begin{table}[!htb]
\centering
\scriptsize
\setlength{\tabcolsep}{2pt}
\begin{tabularx}{\columnwidth}{@{}>{\raggedright\arraybackslash}X r r r r >{\raggedright\arraybackslash}p{0.18\columnwidth}@{}}
\toprule
\textbf{Cat.} & $\bar P_o$ & $\kappa$ & PABAK & Split & Mech. \\
\midrule
Racist & 98.3 & 0.10 & 0.97 & 2.5 & Sparsity \\
Soc. critique & 81.5 & 0.13 & 0.63 & 27.8 & Boundary \\
NIMBY & 93.8 & 0.15 & 0.88 & 9.3 & Sparsity \\
Own opinion & 67.4 & 0.24 & 0.35 & 49.0 & Boundary \\
Solutions & 63.6 & 0.24 & 0.27 & 58.2 & Boundary \\
Fact/claim & 81.3 & 0.26 & 0.63 & 28.7 & Base-rate $\kappa$ \\
Harm. gen. & 90.3 & 0.48 & 0.81 & 18.2 & Sparsity \\
\bottomrule
\end{tabularx}
\caption{Low $\kappa$ mechanisms (category-level means). $\bar P_o$: pairwise agreement; PABAK $=2P_o-1$. Mech.: sparsity, base-rate $\kappa$ paradox, or boundary subjectivity.}
\label{tab:iaa_kappa_mechanisms}
\end{table}

\begin{table}[!htb]
\centering
\scriptsize
\setlength{\tabcolsep}{2pt}
\resizebox{\columnwidth}{!}{%
\begin{tabular}{lrrrr}
\toprule
\textbf{Cat.} & \textbf{Rdt} & \textbf{News} & \textbf{Mtgs} & \textbf{X} \\
\midrule
Genuine Q & 0.77 & 0.77 & 0.79 & 0.68 \\
Rhetorical Q & 0.79 & 0.76 & 0.66 & 0.80 \\
Fact/claim & 0.65 & 0.66 & 0.31 & 0.54 \\
Observation & 0.64 & 0.37 & 0.42 & 0.41 \\
Own opinion & 0.62 & 0.60 & 0.53 & 0.68 \\
Others' op. & 0.57 & 0.43 & 0.22* & 0.47 \\
Money/aid & 0.52 & 0.69 & 0.72 & 0.75 \\
Gov. critique & 0.67 & 0.59 & 0.60 & 0.69 \\
Soc. critique & 0.56 & 0.47 & 0.35 & 0.50 \\
Solutions & 0.63 & 0.59 & 0.53 & 0.63 \\
Personal & 0.74 & 0.45* & 0.53 & 0.58 \\
Media & 0.45 & 0.26* & 0.50* & 0.26 \\
NIMBY & 0.51 & 0.00* & 0.49* & 0.52 \\
Harm. gen. & 0.58 & 0.38* & 0.44 & 0.62 \\
Deservingness & 0.46 & 0.28* & 0.29* & 0.47 \\
Racist & 0.53* & 0.00* & 0.00* & 0.53* \\
\bottomrule
\end{tabular}%
}
\caption{Cohen's $\kappa$ (annotator vs.\ 2-of-3 gold). $^{*}$: $<$5 gold positives.}
\label{tab:humans_vs_gold_kappa}
\end{table}
\FloatBarrier

\begin{table}[!htb]
\centering
\scriptsize
\setlength{\tabcolsep}{2pt}
\begin{tabularx}{\columnwidth}{@{}>{\raggedright\arraybackslash}X rrrr@{}}
\toprule
\textbf{Cat.} & 3/3+ & 2/3+ & 2/3- & 3/3- \\
\midrule
Genuine Q & 46 & 66 & 90 & 1495 \\
Rhetorical Q & 53 & 95 & 75 & 1475 \\
Fact/claim & 1077 & 338 & 171 & 111 \\
Observation & 43 & 105 & 271 & 1279 \\
Own opinion & 418 & 489 & 330 & 461 \\
Others' op. & 15 & 96 & 341 & 1246 \\
Money/aid & 136 & 172 & 378 & 1011 \\
Gov. critique & 55 & 148 & 256 & 1239 \\
Soc. critique & 19 & 133 & 360 & 1186 \\
Solutions & 205 & 472 & 498 & 522 \\
Personal & 28 & 52 & 127 & 1491 \\
Media & 1 & 9 & 105 & 1583 \\
NIMBY & 4 & 38 & 135 & 1521 \\
Harm. gen. & 28 & 89 & 241 & 1339 \\
Deservingness & 2 & 21 & 121 & 1554 \\
Racist & 3 & 2 & 42 & 1651 \\
\midrule
TOTAL & 2133 & 2325 & 3541 & 19164 \\
\bottomrule
\end{tabularx}
\caption{Raw-score breakdown (all sources; $n{=}1{,}698$). Cell counts by category; columns sum to $n$ per row (raw scores 0--3).}
\label{tab:annotator_agreement_breakdown_all_sources}
\end{table}

\FloatBarrier
\section{LLM Prompt}
\label{sec:llm_prompt}
All LLM classification experiments use the same multi-label prompt template; we reproduce it here for replication.
We pass the following prompt into the LLM when classifying the data.
At the beginning of each prompt, we provide definitions of the classification categories. For in-context learning experiments, we additionally include five labeled examples drawn from the gold standard, selected to reflect a range of sources and bias categories. Because this task is multi-label, models may assign multiple labels to a single input. All prompts are independent calls, preventing data leakage between examples.
The template was derived from OATH category wording, expanded with partner-reviewed definitions for the new categories, and frozen after the same pilot used for annotator training ($\approx$20 items/source). We did not run a large automated prompt search. Prompt-order and role-prefix variants, plus decoding hyperparameters, are documented in Appendix~\ref{sec:prompt_sensitivity} and Table~\ref{tab:model_card}.
{\small
\ttfamily
You are an expert in social behavior analysis. Your task is to analyze \{content\_desc\} about homelessness and categorize them according to specific criteria.\\[4pt]

\textbf{DEFINITIONS:}\\
1. Comment Types (select all that apply):\\
- Ask a Genuine Question: The speaker asks a sincere question about homelessness or related issues\\
- Ask a Rhetorical Question: The speaker asks a question not intended to be answered, often to make a point\\
- Provide a Fact or Claim: The speaker provides a factual statement or claim about homelessness\\
- Provide an Observation: The speaker shares an observation about homelessness or related situations\\
- Express Their Opinion: The speaker expresses their own views or feelings about homelessness\\
- Express Others' Opinions: The speaker describes or references the views or feelings of others about homelessness\\[4pt]

2. Critique Categories (select all that apply):\\
- Money Aid Allocation: Discussion of financial resources, aid distribution, or resource allocation for homelessness\\
- Government Critique: Criticism of government policies, laws, or political approaches to homelessness\\
- Societal Critique: Criticism of social norms, systems, or societal attitudes toward homelessness\\[4pt]

3. Response Categories (select all that apply):\\
- Solutions/Interventions: Discussion of specific solutions, interventions, or charitable actions\\[4pt]

4. Perception Types (select all that apply):\\
- Personal Interaction: Direct personal experiences with PEH\\
- Media Portrayal: Discussion of PEH as portrayed in media\\
- Not in my Backyard: Opposition to local homelessness developments\\
- Harmful Generalization: Negative stereotypes about PEH\\
- Deserving/Undeserving: Judgments about who deserves help\\[4pt]

5. Racist Classification:\\
- Yes: Contains explicit or implicit racial bias\\
- No: No racial bias present\\[6pt]

\textbf{INSTRUCTIONS:}\\
1. Read the comment carefully\\
2. Analyze it according to the categories above\\
3. Provide your analysis in the exact format below\\
4. Include a brief reasoning for your classification\\[4pt]

\textbf{FORMAT YOUR RESPONSE EXACTLY AS FOLLOWS:}\\[4pt]
Comment Type: [ask a genuine question, ask a rhetorical question, provide a fact or claim, provide an observation, express their opinion, express others opinions]\\
Critique Category: [money aid allocation, government critique, societal critique]\\
Response Category: [solutions/interventions]\\
Perception Type: [personal interaction, media portrayal, not in my backyard, harmful generalization, deserving/undeserving]\\
Racist: [Yes/No]\\
Reasoning: [brief explanation]
}
If it is zero-shot learning, we just pass in the sentence after. If it is in-context learning, we pass in 5 examples that the human annotators have agreed upon.\\ 

\textbf{Reddit Comments Examples}

\begin{itemize}[leftmargin=*]
    \item ``Are you implying that local police beat panhandlers with batons? Because they don't.'
    \begin{itemize}
        \item  ask a genuine question; provide a fact or claim; societal critique
    \end{itemize}

    \item ``Most comments are saying how great it is to homeless (and it usually is) but are ignoring or unaware of the ***type*** of homeless they plan to [STREET] here. *Drug addicts and people with mental issues.* If it were more homes for homeless and/or low income families, I wouldn't think twice about it but I'm very concerned about a facility housing drug addicts and people with mental issues just a couple hundred feet from a school in the middle of a residential neighborhood.''
    \begin{itemize}
        \item express their opinion; express others opinions; not in my backyard; harmful generalization; deserving/undeserving
    \end{itemize}

    \item ``What is up with the pots and pans? What homeless or trafficked person needs those? Oh wait! She needs some. Send her 50 sets. She can keep one and sell the rest! What a piece of [emoji]''\\
    \begin{itemize}
        \item ask a rhetorical question; express their opinion; societal critique; not in my backyard; harmful generalization
    \end{itemize}

    \item ``I live here too [ORGANIZATION][ORGANIZATION][ORGANIZATION] Fuck the homeless''\\
    \begin{itemize}
        \item express their opinion; not in my backyard; harmful generalization; deserving/undeserving
    \end{itemize}

    \item ``I won't support organizations that are homophobic personally. I clearly stated that others can make their own choices. I then brought up a very real issue in [ORGANIZATION]. I'm a Social Worker. I've worked directly with [ORGANIZATION] in the past. They are very religious. It is what it is. You overreacted to my post IMO. I'm not that important. Its just my opinion. But yeah, I'm not okay with discrimination, so personally I would not work for nor support [PERSON]. I know far too many GLBTQIA+ and Trans individuals that have struggled in [ORGANIZATION] because of discrimination from places like this. Trans houseless individuals in particular are often sexually assaulted around here when they start engaging in services. Its a problem. >""Get over it"" **No.**''
    \begin{itemize}
        \item provide a fact or claim; provide an observation; express their opinion; societal critique
    \end{itemize}
\end{itemize}

\textbf{X (Twitter) Posts Examples}

\begin{itemize}[leftmargin=*]
    \item ``[PERSON] awarded \$100,000 to [PERSON] (ORG3) to enhance employment and education-related skills for [DATE] and migrant farmworkers. The award was part of a \$300,000 discretionary fund award under the CSBG Program. [PERSON]''\\
    provide a fact or claim; money aid allocation; solutions/interventions

    \item ``Did your Black flunky mayor get the[emoji][ORGANIZATION]'s memo 2 stick it 2 Rump instead of serving you by refusing 2 deport migrants + give them Black taxpayers'[emoji]4 shelter+food while Black citizens go homeless? [ORGANIZATION] mayors did. Charity starts at [emoji]. [URL]''
    \begin{itemize}
        \item ask a rhetorical question; provide a fact or claim; express their opinion; money aid allocation; government critique; harmful generalization; deserving/undeserving; racist: Yes
    \end{itemize}

    \item ``PERSON0 Instead of peacocking on social media for your next job, how about you concentrate on the gaggles of homeless people in [ORGANIZATION]?''
    \begin{itemize}
        \item ask a rhetorical question; provide a fact or claim; express their opinion; societal critique; solutions/interventions
    \end{itemize}

    \item ``[ORGANIZATION] Just what [ORGANIZATION] needs...another beggar.''
    \begin{itemize}
        \item  express their opinion; not in my backyard; harmful generalization; deserving/undeserving
    \end{itemize}

    \item ``[ORGANIZATION] area in [ORGANIZATION] is facing a housing crisis. 40\% of people in this area live in poverty, and the city lacks 20,000 affordable housing units. Initiatives like [ORGANIZATION] to fix old housing, but progress depends on securing funding. [URL]''
    \begin{itemize}
        \item  provide a fact or claim; money aid allocation; solutions/interventions
    \end{itemize}
\end{itemize}

\textbf{News Articles Examples}

\begin{itemize}[leftmargin=*]
    \item ``We applaud this important first step to assure the long-term resolution of homelessness.''
    \begin{itemize}
        \item express their opinion; solutions/interventions
    \end{itemize}

    \item ``60 million for programs to support homeless veterans including 20 million for [ORGANIZATION]. The President proposed to eliminate the program.''
    \begin{itemize}
        \item  provide a fact or claim; solutions/interventions
    \end{itemize}

    \item ``[ORGANIZATION] county commissioners on [ORGANIZATION] weighed options for creating a migrant support services center while city emergency managers opened a busing hub, as dozens of migrants remained in homeless conditions [LOCATION].''
    \begin{itemize}
        \item  provide a fact or claim; solutions/interventions
    \end{itemize}

    \item ``About 1 in 3 people who are homeless in [ORGANIZATION] report having a mental illness or a substance use disorder, and the combination of homelessness and substance use or untreated mental illness has led to very public tragedies.''
    \begin{itemize}
        \item  provide a fact or claim; express their opinion
    \end{itemize}

    \item ``I would imagine she is not being delusional about being unsafe on the streets, [ORGANIZATION], executive director of [ORGANIZATION], told [ORGANIZATION]. [PERSON] specializes in treating mentally ill homeless people. Somewhere in all of this is a hook around the fear she has of being unsafe, especially as a woman who is homeless, and that is not uncommon. There should be a real conversation about that, and it could be very useful for figuring out whats going on with her.''
    \begin{itemize}
        \item  provide a fact or claim; provide an observation; express their opinion; solutions/interventions
    \end{itemize}
\end{itemize}

\textbf{Meeting Minutes Examples}

\begin{itemize}[leftmargin=*]
    \item ``Okay. But on that area, one corner is the location that is on dispute right now. And the reason that we are, that I am questioning that is because we got a lot of homeless people there.''
    \begin{itemize}
        \item  provide a fact or claim; express their opinion; not in my backyard
    \end{itemize}

    \item ``but they stuck with us, they got all the permissions they needed, and we would not have made the functional end of veteran homelessness in [ORGANIZATION] without them, so thank you. PERSON0? Well, thank you for this honor.''
    \begin{itemize}
        \item  provide a fact or claim; express their opinion; solutions/interventions
    \end{itemize}

    \item ``I've seen it all, like certain people being removed out of there. And I'm down there [ORGANIZATION]. And all the [ORGANIZATION] and all the residents there, like, I know them all, you know, and I love them because I was dropped off to be homeless [DATE].''
    \begin{itemize}
        \item  provide a fact or claim; provide an observation; express their opinion
    \end{itemize}

    \item ``Yeah. Yeah, I just think that there's a different ROI for chronically homeless folks. And there's a deeper impact, but a more narrow impact.''
    \begin{itemize}
        \item  provide a fact or claim; express their opinion; money aid allocation; solutions/interventions; deserving/undeserving
    \end{itemize}

    \item ``Simple delays, be they [ORGANIZATION] or [LOCATION], have been a feeble solution. Demolitions of up to 370 affordable houses a year valued at \$100 million dwarfs the city's efforts at spending \$20 million to support affordable housing. Can we really achieve affordable housing through demolition?''
    \begin{itemize}
        \item  ask a rhetorical question; provide a fact or claim; express their opinion; money aid allocation; government critique; solutions/interventions
    \end{itemize}
\end{itemize}

\section{Exploratory Corpus Patterns (GPT-4.1 Labels)}
\label{sec:exploratory_corpus}
The main text evaluates LLMs against gold-standard reference labels only. For \emph{exploratory} trends, we apply GPT-4.1 few-shot labels to the full corpus ($\approx$50k texts). Re-labeling the gold-standard set with the same teacher yields mean absolute prevalence drift 9.7-12.1\,pp by source (e.g., Reddit NIMBY 5.2\% reference vs.\ 10.4\% GPT). Because models over-tag NIMBY on gold (Section~\ref{sec:ro3_calibration}), corpus-wide NIMBY rates are likely upward-biased; directional platform comparisons may still inform hypotheses, but magnitudes must not be read as ground-truth prevalence.

\textit{Harmful generalization} and \textit{deserving/undeserving} are more prevalent on Reddit and X than in news or council text (Bonferroni-corrected \citep{weisstein2004bonferroni}). \textit{Harmful generalization} and \textit{solutions/interventions} correlate negatively ($r{=}{-}0.87$; Section~\ref{sec:correlations}). The Negative Bias Frame is elevated for X and news in large-city clusters (Figure~\ref{fig:bias_city_size}). These patterns inherit GPT-4.1 teacher bias and are not core deployment evidence.

\subsection{Reddit Engagement Under Teacher Labels}
\label{sec:reddit_engagement_appendix}
We also tested whether teacher-assigned Negative Bias Frame labels co-occur with Reddit submission score (upvotes minus downvotes). Table~\ref{tab:reddit_engagement} reports medians under GPT-4.1 labels. \textit{Not in my backyard} attains the highest median score ($+2$ vs.\ corpus median 3). Because GPT-4.1 over-tags NIMBY on gold, the NIMBY-tagged set likely mixes opposition posts with mislabeled pro-service housing text; this analysis is illustrative only and should not be cited without prevalence-gap calibration.

\begin{table}[t]
\centering
\footnotesize
\setlength{\tabcolsep}{3pt}
\begin{tabular}{@{}lrrrl@{}}
\toprule
\textbf{Frame} & $n$ & \textbf{Med.} & $\boldsymbol{\Delta}$ & \textbf{Sig.} \\
\midrule
Ask rhetorical question & 3{,}868 & 2 & $-1$ & $p{<}10^{-30}$ \\
Not in my backyard & 3{,}404 & 5 & $+2$ & $p{<}10^{-15}$ \\
Harmful generalization & 10{,}236 & 4 & $+1$ & $p{<}10^{-30}$ \\
Deserving/undeserving & 11{,}998 & 4 & $+1$ & $p{<}10^{-25}$ \\
Racist & 58 & 2 & $-1$ & n.s. \\
\bottomrule
\end{tabular}
\caption{Reddit submission score by Negative Bias Frame (GPT-4.1 labels; exploratory). Med.\ = median among posts with the frame; $\Delta$ = difference from corpus median (3). Bonferroni-corrected Mann-Whitney $U$.}
\label{tab:reddit_engagement}
\end{table}

\section{Category Correlations}
\label{sec:correlations}
Section~\ref{sec:exploratory_corpus} uses GPT-4.1 few-shot labels on the full corpus ($\approx$50k texts); here we report the co-occurrence structure behind exploratory platform comparisons. Pearson $r$ is computed across 38 city-source prevalence vectors (each vector gives the percent of posts in a city $\times$ source cell tagged with a category). Strongest associations include \textit{deserving/undeserving} vs.\ \textit{societal critique} ($r{=}0.84$), \textit{harmful generalization} vs.\ \textit{solutions/interventions} ($r{=}{-}0.87$), and \textit{provide a fact or claim} vs.\ \textit{deserving/undeserving} ($r{=}{-}0.72$). Other significant negative correlations include \textit{solutions/interventions} vs.\ \textit{societal critique} ($r{=}{-}0.75$) and vs.\ \textit{personal interaction} ($r{=}{-}0.75$); significant positive correlations include \textit{express opinion} vs.\ \textit{deserving/undeserving} ($r{=}0.77$) and \textit{ask a genuine question} vs.\ \textit{deserving/undeserving} ($r{=}0.72$). These magnitudes echo the main-text finding that stigma-heavy categories rarely co-occur with constructive shelter framing ($r{=}{-}0.87$ for harmful generalization vs.\ solutions/interventions): when harmful generalizations appear, solutions language is less common, and the link between \textit{deserving/undeserving} and genuine questions may reflect questioning of blame narratives \citep{Sandel2020a,Desmond2023}.

\begin{figure*}[t]
    \centering
    \includegraphics[width=\linewidth]{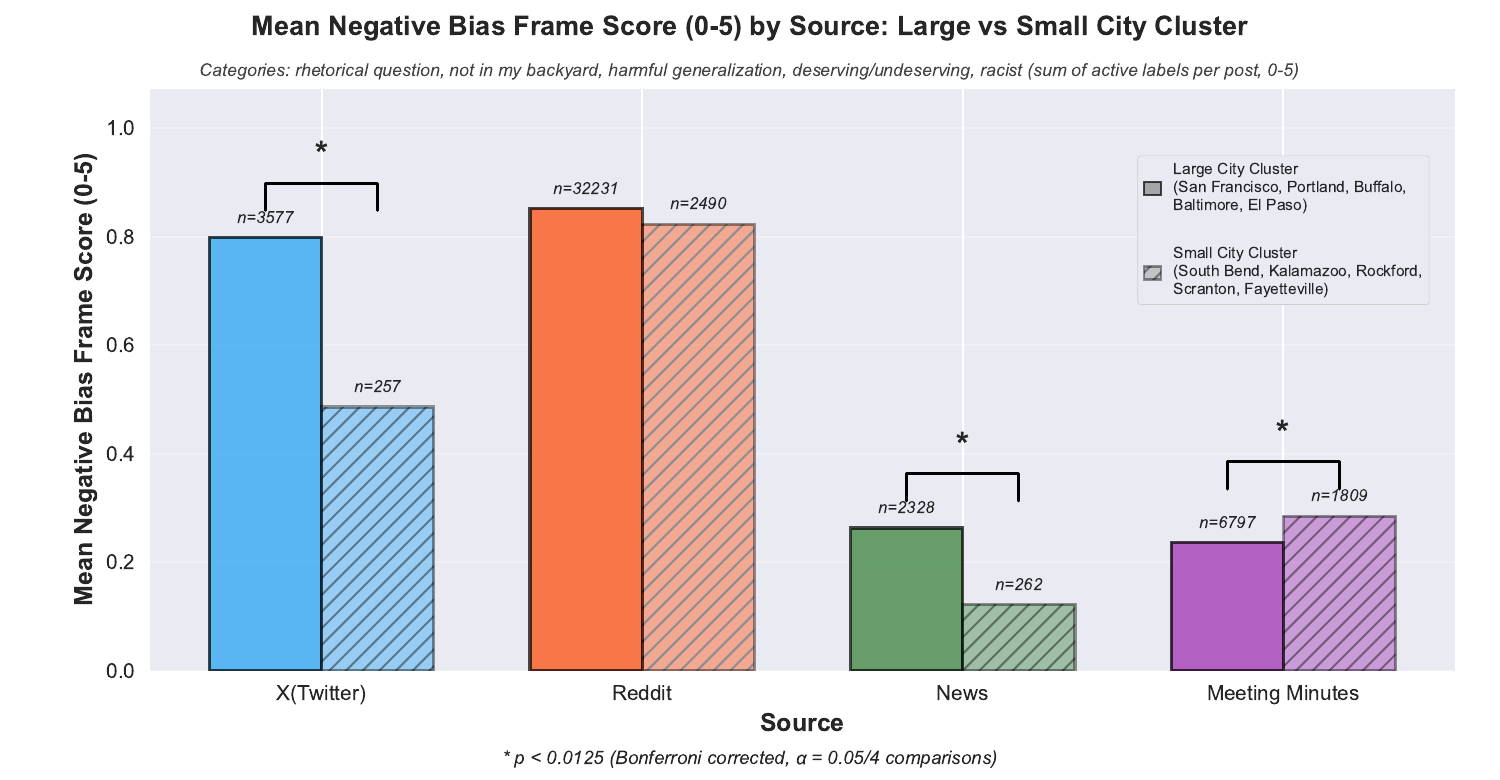}
    \caption{Negative Bias Frame prevalence by city cluster and source (GPT-4.1 labels on full corpus; exploratory). Large and small city clusters each contain five cities (see legend).}
    \label{fig:bias_city_size}
\end{figure*}

Figure~\ref{fig:bias_city_size} plots mean Negative Bias Frame score (0-5 active frame labels) by data source for large- versus small-city clusters (five cities each; legend in figure), using exploratory GPT-4.1 labels. Reddit and X carry higher frame prevalence than news and meeting minutes; \textit{harmful generalization} and \textit{deserving/undeserving} concentrate on open social platforms; negative bias is relatively elevated for X and news in large-city clusters, while council speech stays comparatively constructive.

\section{Annotation Examples}
\label{sec:annotation_examples}
As noted in Section~\ref{sec:ro3_calibration}, multi-label PEH bias annotation is inherently subjective. This appendix provides full gold-standard examples referenced from the main Results section.
To illustrate annotation complexity, we provide two gold-standard items. \\\textbf{X:} ``Did your Black flunky mayor get the[image][ORGANIZATION]'s memo 2 stick it 2 Rump instead of serving you by refusing 2 deport migrants + give them Black taxpayers'[image] 4 shelter+food while Black citizens go homeless? [ORGANIZATION] mayors did. Charity starts at [image].[URL]'' with labels: ask a rhetorical question, provide a fact or claim, express opinion, money aid allocation, harmful generalization, deserving/undeserving, racist. \\\textbf{Meeting minutes:} supportive veteran-housing remarks with labels: provide a fact or claim, express opinion, solutions/interventions.

\section{NIMBY Over-Tagging Error Analysis}
\label{sec:nimby_error_analysis}
Section~\ref{sec:nimby_error_main} summarizes consensus false positives and key trigger rates (Tables~\ref{tab:nimby_triggers_main} and~\ref{tab:calibration_by_model} in the main text). Table~\ref{tab:nimby_cohorts} defines comparison groups on the 1,698-item gold-standard set. Table~\ref{tab:nimby_triggers} lists the full set of lexical and structural cues (regular-expression patterns over post text).

These patterns are \textbf{descriptive co-occurrence}, not causal ablations. The central pattern is asymmetric: housing lexicon and question marks are \emph{more} common among consensus FPs than among true negatives, but opposition verbs are \emph{less} common among FPs than among gold NIMBY positives (8.5\% vs.\ 16.7\%). Affordable-housing phrases are also under-represented among FPs ($-$5.7 pp vs.\ TNs). Together, this supports the account that models tag NIMBY when text discusses housing or asks a question, not when it contains explicit siting opposition.

\begin{table}[t]
\centering
\small
\setlength{\tabcolsep}{3pt}
\begin{tabularx}{\columnwidth}{@{}>{\raggedright\arraybackslash}Xrr@{}}
\toprule
\textbf{Group} & $\boldsymbol{n}$ & \textbf{\%} \\
\midrule
Gold-standard (total) & 1{,}698 & 100.0 \\
Gold NIMBY positive & 42 & 2.5 \\
Consensus FP (gold$-$; $\geq 3$/6 NIMBY) & 117 & 6.9 \\
Any-model FP (gold$-$; $\geq 1$/6 NIMBY) & 988 & 58.1 \\
True negative (gold$-$; 0/6 NIMBY) & 672 & 39.6 \\
\bottomrule
\end{tabularx}
\caption[NIMBY error-analysis cohorts]{NIMBY error-analysis cohorts on the gold-standard set (six zero-shot prompt LLMs). FP = false positive; consensus FP requires $\geq 3$ model NIMBY tags among gold-negative items.}
\label{tab:nimby_cohorts}
\end{table}

\begin{table*}[t]
\centering
\footnotesize
\setlength{\tabcolsep}{5pt}
\begin{tabular}{@{}lrrrr@{}}
\toprule
\textbf{Trigger} & \textbf{FP (\%)} & \textbf{TN (\%)} & \textbf{Gold NIMBY$+$ (\%)} & $\boldsymbol{\Delta}$ \textbf{(pp)} \\
\midrule
Question mark & 32.5 & 9.7 & 31.0 & +22.8 \\
Housing / homeless lexicon & 89.7 & 72.2 & 92.9 & +17.6 \\
Rhetorical cue (e.g., ``did you read'') & 6.8 & 0.1 & 4.8 & +6.7 \\
Proximity / local & 9.4 & 3.9 & 21.4 & +5.5 \\
Opposition verbs & 8.5 & 4.0 & 16.7 & +4.5 \\
Development / zoning & 10.3 & 7.6 & 11.9 & +2.7 \\
Explicit ``NIMBY'' / backyard & 2.6 & 0.0 & 9.5 & +2.6 \\
Affordable housing phrase & 11.1 & 16.8 & 11.9 & $-$5.7 \\
Shelter siting & 4.3 & 5.8 & 2.4 & $-$1.5 \\
Policy numbers (\%, units) & 4.3 & 5.5 & 2.4 & $-$1.2 \\
\bottomrule
\end{tabular}
\caption{Trigger rates by group: consensus NIMBY false positives (FP, $n{=}117$), true negatives (TN, $n{=}672$), and gold NIMBY positives ($n{=}42$). $\Delta$ = FP minus TN (percentage points). Rows sorted by $\Delta$.}
\label{tab:nimby_triggers}
\end{table*}

\section{Beyond Lexical Triggers: Representation Probes}
\label{sec:nimby_representation}
Section~\ref{sec:nimby_error_main} shows lexical asymmetries among consensus FPs. Because closed LLM APIs do not expose internals, we probe whether over-tagged texts occupy a distinctive region in frozen TF--IDF+SVD(128) embeddings and whether rule-based edits shift a consensus-FP linear probe (Tables~\ref{tab:nimby_representation},~\ref{tab:nimby_edit_sensitivity}). This complements lexical co-occurrence; it is not an in-model ablation.

\begin{table}[t]
\centering
\small
\begin{tabular}{@{}lrr@{}}
\toprule
\textbf{Centroid pair} & \textbf{Cosine} & $n$ \\
\midrule
FP vs TN & 0.920 & 117/1543 \\
FP vs TP & 0.935 & 117/42 \\
TN vs TP & 0.860 & 1543/42 \\
\midrule
Probe AUC (gold NIMBY) & \multicolumn{2}{l}{0.705} \\
Probe AUC (consensus FP) & \multicolumn{2}{l}{0.709} \\
Probe AUC (FP vs TN $|$ gold-) & \multicolumn{2}{l}{0.728} \\
\bottomrule
\end{tabular}
\caption{TF--IDF+SVD(128) centroid cosines and linear-probe AUC for gold NIMBY vs.\ consensus FPs ($\geq 3/6$). Edit sensitivity: Table~\ref{tab:nimby_edit_sensitivity}.}
\label{tab:nimby_representation}
\end{table}

\begin{table}[!htb]
\centering
\scriptsize
\setlength{\tabcolsep}{2pt}
\begin{tabularx}{\columnwidth}{@{}>{\raggedright\arraybackslash}X r r r r@{}}
\toprule
\textbf{Edit} & $n$ & $\Delta$ & $\downarrow$ & $\uparrow$ \\
\midrule
Strip ``?'' & 38 & +0.000 & 0\% & 0\% \\
Drop housing lex. & 105 & $-0.077$ & 74\% & 3\% \\
Neutralize opp. & 9 & +0.015 & 0\% & 33\% \\
Add opposition & 117 & $-0.040$ & 62\% & 0\% \\
Far proximity & 7 & +0.079 & 0\% & 71\% \\
\bottomrule
\end{tabularx}
\caption{Counterfactual edits on consensus NIMBY FPs ($n{=}117$). $\Delta$: mean probe-score shift (FP-trained, not gold); $\downarrow$/$\uparrow$: share with lower/higher score.}
\label{tab:nimby_edit_sensitivity}
\end{table}

\section{Prevalence-Gap Uncertainty}
\label{sec:prevalence_uncertainty}
Table~\ref{tab:prevalence_gap_ci} reports bootstrap 95\% confidence intervals for pooled prevalence gaps ($\hat{\pi}_{\mathrm{model}}-\pi_{\mathrm{ref}}$), resampling gold-standard items ($n{=}1{,}698$) with model predictions pooled over six prompt LLMs and zero/few-shot. Zero-shot-only CIs, source stratification, and vote-threshold sensitivity appear in Appendix~\ref{sec:zeroshot_calibration_audit}. Intervals are tight for high-prevalence categories because $n$ is large, but \textbf{slice-level} gaps remain unstable when gold positives are sparse (e.g., NIMBY: 42 positives total; 0 in news; deserving/undeserving: 26).

\begin{table}[t]
\centering
\small
\setlength{\tabcolsep}{3pt}
\begin{tabularx}{\columnwidth}{@{}>{\raggedright\arraybackslash}X r c r@{}}
\toprule
\textbf{Category} & \textbf{Gap} & \textbf{95\% CI} & $\boldsymbol{n}_{+}$ \\
\midrule
Not in my backyard & +11.5 & [10.6, 12.6] & 42 \\
Provide fact/claim & $-$30.5 & [$-$31.3, $-$27.4] & 1{,}420 \\
Harmful generalization & +11.1 & [9.9, 12.8] & 118 \\
Express opinion & +3.8 & [0.2, 5.3] & 907 \\
\bottomrule
\end{tabularx}
\caption[Prevalence-gap bootstrap CIs]{Bootstrap 95\% CIs for pooled prevalence gaps ($n{=}1{,}698$ gold-standard items). Gap in percentage points; $n_{+}$: gold positives under 2-of-3 majority ($s_{ij}\geq\frac{2}{3}$).}
\label{tab:prevalence_gap_ci}
\end{table}

\section{Prompt Calibration Audit (Zero-shot vs.\ Few-shot)}
\label{sec:zeroshot_calibration_audit}
Section~\ref{sec:ro3_calibration} and Table~\ref{tab:calibration_by_model} summarize prevalence gaps pooled across prompt modes. Here we report a parallel audit on gold-subset flags for \textbf{zero-shot} and \textbf{few-shot} separately (LLaMA, Phi-4, Qwen, GPT-4.1, Gemini, Grok). \textbf{Human reference} labels use three annotators ($s_{ij}\in\{0,\frac{1}{3},\frac{2}{3},1\}$; default 2-of-3, Eq.~1). \textbf{Model aggregation} uses six judges within a prompt mode (vote fraction $v$). Because the NIMBY error analysis (Section~\ref{sec:nimby_error_main}) and consensus false positives are defined on \textbf{six zero-shot} prompt LLMs, we emphasize zero-shot results when connecting to that error analysis.

\paragraph{Does few-shot change calibration?}
Using six-model vote aggregation at $v\geq 0.5$, few-shot reduces NIMBY inflation (+5.7 pp $\rightarrow$ +2.8 pp) and reduces fact/claim under-detection ($-$20.7 pp $\rightarrow$ $-$7.1 pp), but increases over-tagging of express opinion (+13.1 pp $\rightarrow$ +19.5 pp) and harmful generalization (+4.3 pp $\rightarrow$ +6.8 pp). Thus few-shot can shift calibration substantially, so prompt mode should be treated as part of the monitor specification rather than an implementation detail.

\begin{table*}[t]
\centering
\small
\setlength{\tabcolsep}{3pt}
\begin{tabular}{@{}l r c r@{}}
\toprule
\textbf{Category} & \textbf{Zero-shot} & \textbf{Few-shot} & $n_{+}$ \\
\midrule
Not in my backyard & +5.7 [4.4, 7.0] & +2.8 [1.8, 3.8] & 46 / 42 \\
Provide fact/claim & $-$20.7 [$-$23.2, $-$18.2] & $-$7.1 [$-$9.3, $-$4.6] & 1{,}423 / 1{,}420 \\
Express opinion & +13.1 [10.7, 15.5] & +19.5 [17.0, 21.9] & 910 / 907 \\
Harmful generalization & +4.3 [2.6, 5.9] & +6.8 [5.1, 8.5] & 122 / 118 \\
\bottomrule
\end{tabular}
\caption{Pooled prevalence gaps (pp) with bootstrap 95\% CIs, by prompt mode (six models; all sources; $v\geq 0.5$; gold 2-of-3). $n_{+}$ reports gold positives (zero-shot / few-shot) because model coverage differs slightly by prompt mode.}
\label{tab:gap_zero_vs_few}
\end{table*}

\paragraph{Per-model prevalence gaps (zero-shot).}
Table~\ref{tab:calibration_by_model_zeroshot} gives $\hat{\pi}-\pi_{\mathrm{ref}}$ (pp) per model under zero-shot, pooled over four sources. Gold $\pi_{\mathrm{ref}}$ uses \textbf{2-of-3} annotator majority ($s_{ij}\geq\frac{2}{3}$, Eq.~1). Qwen remains the largest NIMBY inflator ($+37.5$ pp); GPT-4.1 and Grok are smallest among closed APIs ($+4.3$, $+1.7$ pp). Few-shot differences are summarized above at the pooled level.

\begin{table*}[!tbp]
\centering
\footnotesize
\setlength{\tabcolsep}{3pt}
\begin{tabular}{@{}lrrrrrrr@{}}
\toprule
\textbf{Category} & \textbf{LLaMA} & \textbf{Phi-4} & \textbf{Qwen} & \textbf{GPT-4.1} & \textbf{Gemini} & \textbf{Grok} & \textbf{Mean} \\
\midrule
Not in my backyard & +15.7 & +16.6 & +37.5 & +4.3 & +5.2 & +1.7 & +13.5 \\
Provide fact/claim & $-$55.2 & $-$73.6 & $-$28.7 & $-$6.5 & $-$0.8 & $-$39.0 & $-$33.9 \\
Express opinion & +12.5 & $-$22.6 & +29.8 & +7.7 & +6.0 & $-$26.2 & +1.2 \\
Harmful generalization & +9.4 & $-$2.4 & +37.5 & +12.0 & +10.6 & $-$2.0 & +10.9 \\
\bottomrule
\end{tabular}
\caption{Prevalence gap (pp) by zero-shot model on the gold-standard set ($\hat{\pi}_{\mathrm{model}}-\pi_{\mathrm{ref}}$; all sources pooled). \textbf{Mean} = unweighted average of six models. Positive = over-tagging.}
\label{tab:calibration_by_model_zeroshot}
\end{table*}

\paragraph{Pooled gaps, bootstrap CIs, and vote aggregation.}
For deployment, municipalities may aggregate multiple model tags. We define the \textbf{vote fraction} $v \in \{0,\frac{1}{6},\ldots,1\}$ as the fraction of zero-shot models assigning a label. Table~\ref{tab:prevalence_gap_ci_zeroshot} reports pooled gaps when binarizing at $v\geq 0.5$ (majority of six), with bootstrap 95\% CIs (2{,}000 resamples). NIMBY inflation is $+5.7$ pp [4.4, 7.0]; fact/claim under-detection is $-$20.7 pp [$-$23.2, $-$18.2]---smaller in magnitude than the zero/few-shot pooled main-text table but the same direction.

Gaps are \textbf{not invariant} to the vote threshold (Table~\ref{tab:vote_threshold_sweep}). For NIMBY, requiring any model to tag ($v\geq\frac{1}{6}$) yields $+21.0$ pp inflation vs.\ gold; majority vote ($v\geq 0.5$) yields $+5.7$ pp; unanimous agreement ($v{=}1$) yields $-$2.6 pp (under-tagging). Fact/claim under-detection worsens from $-$2.9 pp (any-model) to $-$80.9 pp (unanimous). Partners should treat the aggregation rule as part of the monitor specification, not a neutral default.

\begin{table}[t]
\centering
\small
\setlength{\tabcolsep}{3pt}
\begin{tabularx}{\columnwidth}{@{}>{\raggedright\arraybackslash}X r c r@{}}
\toprule
\textbf{Category} & \textbf{Gap} & \textbf{95\% CI} & $n_{+}$ \\
\midrule
Not in my backyard & +5.7 & [4.4, 7.0] & 46 \\
Provide fact/claim & $-$20.7 & [$-$23.2, $-$18.2] & 1{,}423 \\
Express opinion & +13.1 & [10.7, 15.5] & 910 \\
Harmful generalization & +4.3 & [2.6, 5.9] & 122 \\
\bottomrule
\end{tabularx}
\caption{Bootstrap 95\% CIs for pooled zero-shot prevalence gaps ($n{=}1{,}708$; gold 2-of-3). Gap at six-model vote $v\geq 0.5$.}
\label{tab:prevalence_gap_ci_zeroshot}
\end{table}

\begin{table}[t]
\centering
\small
\setlength{\tabcolsep}{4pt}
\begin{tabular}{@{}l r r r@{}}
\toprule
\textbf{Vote rule} & \textbf{NIMBY} & \textbf{Fact/claim} & \textbf{Expr.\ op.} \\
\midrule
Any model ($v\geq\frac{1}{6}$) & +21.0 & $-$2.9 & +30.6 \\
Majority ($v\geq 0.5$) & +5.7 & $-$20.7 & +13.1 \\
All six ($v{=}1$) & $-$2.6 & $-$80.9 & $-$45.7 \\
\bottomrule
\end{tabular}
\caption{Prevalence gap (pp) vs.\ gold at three vote-aggregation rules (six zero-shot models).}
\label{tab:vote_threshold_sweep}
\end{table}

\paragraph{Source-wise gaps.}
Table~\ref{tab:prevalence_gap_by_source} stratifies gaps by modality (gold 2-of-3; six-model vote $v\geq 0.5$). Council minutes show the largest fact/claim under-detection ($-$33.0 pp); X is similar ($-$40.4 pp). News has \textbf{zero} gold NIMBY positives under 2-of-3 ($n{=}382$); the $+2.1$ pp gap is unstable (marked $^*$). Reddit carries most gold NIMBY support ($n_{+}{=}30$) and the largest NIMBY inflation ($+10.9$ pp). Slice-level gaps should not be read as precise point estimates when $n_{+}<5$.

\begin{table*}[t]
\centering
\footnotesize
\setlength{\tabcolsep}{3pt}
\begin{tabular}{@{}l l r c r@{}}
\toprule
\textbf{Source} & \textbf{Category} & \textbf{Gap} & \textbf{95\% CI} & $n_{+}$ \\
\midrule
Reddit & Not in my backyard & +10.9 & [7.7, 14.0] & 30 \\
Reddit & Provide fact/claim & $-$5.9 & [$-$10.7, $-$0.8] & 285 \\
News & Not in my backyard & +2.1$^*$ & [0.8, 3.9] & 0 \\
News & Provide fact/claim & $-$5.2 & [$-$8.9, $-$1.6] & 357 \\
Meeting minutes & Provide fact/claim & $-$33.0 & [$-$38.2, $-$27.7] & 351 \\
X (Twitter) & Provide fact/claim & $-$40.4 & [$-$45.4, $-$35.5] & 430 \\
\midrule
\multicolumn{5}{@{}l@{}}{\footnotesize ($^*$$<5$ gold positives; interpret slice-level gaps with caution.)} \\
\bottomrule
\end{tabular}
\caption{Selected source-wise prevalence gaps (pp; gold 2-of-3; six zero-shot models, $v\geq 0.5$).}
\label{tab:prevalence_gap_by_source}
\end{table*}

\paragraph{Gold reference: three annotators, two vote cuts.}
Each item has exactly three human coders. The \textbf{soft score} $s_{ij}$ is the fraction of coders who label category $j$ positive on item $i$, so $s_{ij}\in\{0,\tfrac{1}{3},\tfrac{2}{3},1\}$ (0, 1, 2, or 3 of 3). Only two non-trivial binarizations arise: \textbf{1-of-3} ($s_{ij}\geq\tfrac{1}{3}$; any coder positive) and \textbf{2-of-3} ($s_{ij}\geq\tfrac{2}{3}$; majority vote, Section~\ref{sec:bias_classification}). Main text and Table~\ref{tab:calibration_by_model_zeroshot} use \textbf{2-of-3}. Table~\ref{tab:gold_tau_sensitivity} shows that LLM--gold gaps are sensitive to this choice when annotators split: on Reddit fact/claim, the gap is $-30.2$\,pp under 1-of-3 but $-5.9$\,pp under 2-of-3; NIMBY shifts from $-5.3$\,pp (1-of-3) to $+10.9$\,pp (2-of-3). We do not apply a generic fractional cutoff such as $0.5$ to $s_{ij}$: with three annotators only the vote counts $\{0,1,2,3\}$ matter, and requiring $s_{ij}\geq 0.5$ is the same as 2-of-3 because $\tfrac{2}{3}>\tfrac{1}{2}$.

\begin{table}[t]
\centering
\small
\setlength{\tabcolsep}{3pt}
\begin{tabular}{@{}l l r r@{}}
\toprule
\textbf{Source} & \textbf{Category} & \textbf{1-of-3} & \textbf{2-of-3} \\
 & & ($s_{ij}\geq\tfrac{1}{3}$) & ($s_{ij}\geq\tfrac{2}{3}$) \\
\midrule
Reddit & NIMBY & $-$5.3 & $+$10.9 \\
Reddit & Fact/claim & $-$30.2 & $-$5.9 \\
News & Fact/claim & $-$9.7 & $-$5.2 \\
Meeting minutes & Fact/claim & $-$36.0 & $-$33.0 \\
X (Twitter) & Fact/claim & $-$44.7 & $-$40.4 \\
\bottomrule
\end{tabular}
\caption{Prevalence gap (pp) under 1-of-3 vs.\ 2-of-3 gold (six zero-shot models, $v\geq 0.5$). \textbf{2-of-3} is the paper's reference rule.}
\label{tab:gold_tau_sensitivity}
\end{table}

\paragraph{Reliability and expected calibration error (ECE).}
Prompt LLMs return binary flags, not probabilities. We treat vote fraction $v$ as a proxy confidence and compute ECE (10 equal-width bins) against \textbf{2-of-3} gold. Table~\ref{tab:ece_zeroshot} summarizes ECE by category; Table~\ref{tab:reliability_bins} lists illustrative high-support bins ($n\geq 20$). Fact/claim has the highest ECE (0.34). Express opinion is best calibrated overall (ECE 0.04). For NIMBY, when $v\in[0.5,0.6)$ ($n{=}97$), only 13.4\% of items are gold-positive despite $\bar{v}=0.50$---models assign moderate consensus tags where gold rarely agrees. For fact/claim at the same bin ($n{=}479$), $\pi_{\mathrm{gold}}=87.5\%$ exceeds $\bar{v}=0.50$, reflecting systematic under-tagging relative to human majority.

\begin{table}[t]
\centering
\small
\begin{tabular}{@{}l r r@{}}
\toprule
\textbf{Category} & \textbf{ECE} & $n_{+}$ \\
\midrule
Provide fact/claim & 0.342 & 1{,}416 \\
Not in my backyard & 0.138 & 42 \\
Harmful generalization & 0.113 & 118 \\
Express opinion & 0.037 & 907 \\
\bottomrule
\end{tabular}
\caption{ECE for vote fraction vs.\ 2-of-3 gold (six zero-shot models; $n{=}1{,}698$).}
\label{tab:ece_zeroshot}
\end{table}

\begin{table}[t]
\centering
\footnotesize
\setlength{\tabcolsep}{3pt}
\begin{tabular}{@{}l c r r r@{}}
\toprule
\textbf{Category} & \textbf{$v$ bin} & $n$ & $\bar{v}$ & $\pi_{\mathrm{gold}}$ \\
\midrule
NIMBY & [0.0, 0.1) & 678 & 0.00 & 0.6\% \\
NIMBY & [0.5, 0.6) & 97 & 0.50 & 13.4\% \\
Fact/claim & [0.3, 0.4) & 404 & 0.33 & 79.2\% \\
Fact/claim & [0.5, 0.6) & 479 & 0.50 & 87.5\% \\
Fact/claim & [0.8, 0.9) & 201 & 0.83 & 96.0\% \\
\bottomrule
\end{tabular}
\caption{Selected reliability bins (six zero-shot models vs.\ 2-of-3 gold; $n\geq 20$). $\bar{v}$: mean vote fraction in bin; $\pi_{\mathrm{gold}}$: gold positive rate.}
\label{tab:reliability_bins}
\end{table}

\paragraph{Lexical counterfactuals and representation probes.}
Table~\ref{tab:nimby_edit_sensitivity} applies rule-based edits to consensus FPs and measures shifts in an FP-trained embedding probe. Housing-lexicon removal lowers scores on 74\% of edits ($\Delta{=}{-}0.077$); question stripping does not; opposition neutralization applies to $n{=}9$ texts. Representation pilots only---not re-queries of the six LLM judges.

\section{Mitigating NIMBY Over-Tagging (Vote Threshold)}
\label{sec:nimby_mitigation}
Reviewers asked whether prevalence gaps can be reduced rather than only diagnosed. Table~\ref{tab:nimby_vote_mitigation} retunes the six-model vote threshold $v$ on gold reference labels. Raising $v$ from $0.5$ to $2/3$ cuts absolute zero-shot NIMBY gap from $5.7$ to $1.9$\,pp. This is a monitor-spec change, not a fix for shallow lexical proxies (Section~\ref{sec:nimby_error_main}). Encoder/LoRA mitigation: Appendix~\ref{sec:evaluation_framing}.

\begin{table}[t]
\centering
\footnotesize
\setlength{\tabcolsep}{3pt}
\begin{tabularx}{\columnwidth}{@{}>{\raggedright\arraybackslash}X rr@{}}
\toprule
\textbf{Setting} & \textbf{Gap} & \textbf{F1} \\
\midrule
\multicolumn{3}{@{}l@{}}{\textit{Baseline}} \\
Pooled per-model (Tab.~\ref{tab:calibration_by_model}) & $+11.5$ & --- \\
\multicolumn{3}{@{}l@{}}{\textit{Zero-shot}} \\
$v\geq\frac{1}{6}$ & $+21.0$ & 0.156 \\
$v\geq\frac{1}{3}$ & $+21.0$ & 0.156 \\
$v\geq\frac{1}{2}$ & $+5.7$ & 0.284 \\
$v\geq\frac{2}{3}$ & $-1.9$ & 0.233 \\
$v\geq\frac{5}{6}$ & $-1.9$ & 0.233 \\
$v=1$ & $-2.6$ & 0.083 \\
\multicolumn{3}{@{}l@{}}{\textit{Few-shot}} \\
$v\geq\frac{1}{6}$ & $+15.2$ & 0.198 \\
$v\geq\frac{1}{3}$ & $+15.2$ & 0.198 \\
$v\geq\frac{1}{2}$ & $+2.8$ & 0.351 \\
$v\geq\frac{2}{3}$ & $-2.1$ & 0.163 \\
$v\geq\frac{5}{6}$ & $-2.1$ & 0.163 \\
$v=1$ & $-2.5$ & 0.000 \\
\bottomrule
\end{tabularx}
\caption{NIMBY gap vs.\ six-model vote threshold $v$ (gold 2-of-3). Zero-shot NIMBY gap falls from $5.7$\,pp at $v\geq\frac{1}{2}$ to $1.9$\,pp at $v\geq\frac{2}{3}$.}
\label{tab:nimby_vote_mitigation}
\end{table}

\section{Prompt Engineering and Sensitivity}
\label{sec:prompt_sensitivity}
The classification prompt in Appendix~\ref{sec:llm_prompt} was derived from OATH frame definitions~\citep{ranjit2024oath}, expanded to our 16-category schema with partner review, and frozen after pilot runs ($\approx$20 items/source). We define three audit variants: \textbf{A} baseline, \textbf{B} within-block shuffle (seed $7$), and \textbf{C} role line removed.

\paragraph{NIMBY prompt sensitivity (local Qwen).}
Table~\ref{tab:nimby_prompt_sensitivity} reports NIMBY gaps on the gold-standard set ($n{=}1{,}698$) with local Qwen2.5-7B-Instruct (MLX 4-bit; $T{=}0.1$). Variant~\textbf{A} reuses stored zero-shot flags; \textbf{B}/\textbf{C} re-infer one category at a time. Gaps stay positive ($+37.6$, $+10.2$, $+39.9$\,pp); shuffle (B) cuts magnitude more than removing the role line (C).

\begin{table}[!htb]
\centering
\scriptsize
\setlength{\tabcolsep}{2pt}
\begin{tabularx}{\columnwidth}{@{}>{\raggedright\arraybackslash}X r r r@{}}
\toprule
\textbf{Variant} & $\pi_{\mathrm{ref}}$ & $\pi_{\mathrm{m}}$ & \textbf{Gap} \\
\midrule
A (stored) & 2.5 & 40.0 & $+37.6$ \\
B (shuffle) & 2.5 & 12.7 & $+10.2$ \\
C (no role) & 2.5 & 42.4 & $+39.9$ \\
\bottomrule
\end{tabularx}
\caption{NIMBY gaps under prompt variants ($n{=}1{,}698$; pp). A: stored Qwen zero-shot; B/C: local MLX re-inference. $\pi_{\mathrm{m}}$: model prevalence.}
\label{tab:nimby_prompt_sensitivity}
\end{table}

\paragraph{Decoding reproducibility.}
Main-text flags use $T{=}0.1$ (Table~\ref{tab:model_card}). We treat gold $\tau$, vote $v$, and prompt templates as primary sensitivity knobs (Tables~\ref{tab:gold_tau_sensitivity},~\ref{tab:nimby_vote_mitigation}); re-audit if decode settings change.

\begin{table}[t]
\centering
\footnotesize
\setlength{\tabcolsep}{3pt}
\begin{tabularx}{\columnwidth}{@{}l l >{\raggedright\arraybackslash}X r@{}}
\toprule
\textbf{Model} & \textbf{Be.} & \textbf{ID} & $T$ \\
\midrule
GPT & API & \texttt{gpt-4.1} & 0.1 \\
Gemini & API & \texttt{gemini-2.5-pro} & 0.1 \\
Grok & API & \texttt{grok-4-latest} & 0.1 \\
LLaMA & local & \texttt{Llama-3.2-3B-Instruct} & 0.1 \\
Qwen & local & \texttt{Qwen2.5-7B-Instruct} & 0.1 \\
Phi-4 & local & \texttt{Phi-4-mini-instruct} & 0.1 \\
\bottomrule
\end{tabularx}
\caption{Model IDs and decode settings ($t_{\max}{=}500$, $T{=}0.1$; local: top-$p{=}0.95$, rep.\ penalty $1.1$). \textbf{Be.}: API or local HF. All models inferred May--Aug 2025.}
\label{tab:model_card}
\end{table}

\section{Municipal Deployment and Audit Sampling}
\label{sec:municipal_deployment}
South Bend partners plan to adopt the monitor after calibration fixes; during development, monthly meetings with [Person B] and nonprofit staff informed taxonomy design and audit thresholds. Intended use is \emph{directional} monitoring (which frames spike after a council vote), not autonomous enforcement. We do not claim a universal F1 floor; we require (i) recurring gold-standard audits on partner-priority categories and (ii) prevalence-gap caps on deployed models before public-facing campaigns (e.g., countering NIMBY narratives).

\textbf{Standing audit sample.} Stratify by source and city as in corpus construction. For a category with reference prevalence $\pi_{\mathrm{ref}}\approx 5\%$ (e.g., NIMBY on Reddit), auditing $n\approx 200$ stratified items per review cycle yields $\sim$10 expected positives (rule-of-thumb for stable $\pi_{\mathrm{ref}}$ estimates within $\pm 5$ pp at 95\% confidence). For rarer labels (deserving/undeserving, racist), audit \textbf{all available positives} in the sample plus matched negatives rather than relying on pooled gaps alone.

\textbf{Frequency.} Re-audit when changing model, prompt, or taxonomy (always), and at least quarterly when labels drive public dashboards. Corpus-wide GPT-4.1 passes remain exploratory between audits.

\section{Teacher Model: GPT-4.1 vs.\ Gemini}
\label{sec:teacher_model}
Gemini 2.5 Pro leads family-average macro-F1 on gold (44.47 vs.\ GPT-4.1 43.45; Table~\ref{tab:llm_benchmark}). We standardize on \textbf{GPT-4.1 few-shot} for corpus labeling because it is within $\approx$1 macro-F1 point of Gemini, our prevalence-gap pipeline uses a single teacher, and both APIs were stable during May--Aug 2025 model inference. Do not mix teachers on one dashboard without re-calibration.

\section{Stakeholder Collaboration and Impact}
\label{sec:stakeholder_collaboration}
This project was motivated by an immediate need in South Bend, Indiana: public backlash over where to locate a new homelessness shelter and amplification of ``not in my backyard'' narratives. We partnered with two South Bend non-profit organizations (Motels4Now and Broadway Christian
Parish) whose staff, including people with lived experience of housing instability, guided our taxonomy (Section~\ref{sec:bias_classification}) and ensured labels reflected real concerns, including tracking \textit{racist} rhetoric separately from other stigmatizing frames.

Throughout the study, monthly meetings with Marshall Smith and partner staff reviewed city- and source-specific findings and shaped the Negative Bias Frame for shelter-siting monitoring. Partners are preparing live dashboards and an anti-NIMBYism campaign aligned with the NIMBY calibration findings in Section~\ref{sec:ro3_calibration}. Partners emphasized that inflated NIMBY rates in LLM monitors could falsely suggest community opposition when discourse is pro-service. We recommend recurring gold-standard audit samples before acting on corpus dashboards. The released corpus, gold-standard annotations, and scripts support municipalities needing repeatable stigma signals (SDG 1: No Poverty; SDG 11: Sustainable Cities), not one-off social-media reads.

\section{Full Prevalence-Gap Tables (Gold-Standard Set)}
\label{sec:calibration_gaps_appendix}
Section~\ref{sec:ro3_calibration} reports per-model prevalence gaps (Table~\ref{tab:calibration_by_model}) in the main text. Table~\ref{tab:underdetection_delta_pooled} below lists all 15 non-\textit{racist} categories, including \textit{express opinion} and \textit{harmful generalization}.

\begin{table*}[t]
\centering
\small
\setlength{\tabcolsep}{3pt}
\resizebox{\textwidth}{!}{%
\begin{tabular}{lccccccc}
\toprule
Category & LLaMA & Phi-4 & Qwen & Gemini & Grok & GPT & Overall \\
\midrule
Ask Genuine Question & 0 & -4.9 & 0 & 0 & -2.9 & +1.2 & -1.2 \\
Ask Rhetorical Question & -4.1 & -7.2 & -2.8 & +3.3 & -2.0 & 0 & -2.1 \\
Provide Fact/Claim & -55.1 & -61.9 & -23.2 & -1.2 & -34.3 & -7.6 & -30.5 \\
Provide Observation & 0 & -4.9 & +36.7 & +22.1 & +18.2 & +36.8 & +18.1 \\
Express Opinion & +16.4 & -16.4 & +25.2 & +7.0 & -20.2 & +10.7 & +3.8 \\
Express Others' Opinions & +2.4 & -5.5 & +14.4 & +12.0 & +2.4 & +10.5 & +6.0 \\
Money/Aid Allocation & -10.1 & -15.2 & +6.0 & +13.2 & -6.7 & +2.0 & -1.8 \\
Government Critique & +19.5 & +9.2 & +26.9 & +23.0 & -1.6 & +11.7 & +14.8 \\
Societal Critique & +21.8 & +3.8 & +16.3 & +17.8 & +2.2 & +27.6 & +14.9 \\
Solutions/Interventions & -5.8 & -19.9 & +3.3 & +19.0 & -6.9 & +4.2 & -1.0 \\
Personal Interaction & -1.6 & -2.9 & +19.2 & +9.5 & +1.5 & +11.0 & +6.1 \\
Media Portrayal & +11.0 & +17.8 & +22.6 & +4.7 & +1.2 & +9.2 & +11.1 \\
Not in My Backyard & +15.9 & +14.4 & +27.4 & +4.8 & +2.4 & +4.0 & +11.5 \\
Harmful Generalization & +20.1 & 0 & +27.2 & +9.8 & 0 & +9.8 & +11.1 \\
Deserving/Undeserving & +20.5 & +4.5 & +20.1 & +12.0 & +2.6 & +17.8 & +12.9 \\
\midrule
\textit{Overall Avg} & +3.4 & -6.0 & +14.6 & +10.4 & -2.9 & +9.9 & +4.9 \\
\bottomrule
\end{tabular}%
}
\caption{Prevalence gaps vs.\ gold reference labels (six prompt LLMs, gold-standard set, pooled). Gap $\hat{\pi}_{\mathrm{model}}-\pi_{\mathrm{ref}}$ (pp), pooled across prompt modes and four sources. Gold uses 2-of-3 majority ($s_{ij}\geq\frac{2}{3}$); negative = under-detection. \textit{Racist} omitted ($<$5 positive instances on several slices).}
\label{tab:underdetection_delta_pooled}
\end{table*}

\FloatBarrier

\section{City Classification}
\label{sec:city_classification}
Section~\ref{sec:exploratory_corpus} compares negative-bias prevalence across city clusters and sources under GPT-4.1 labels.

Even though there is not a significant difference between the group of cities; we can examine the heatmap by city as shown in Figure \ref{fig:city_heatmap}. The intensity of coloring represents the percentage of content that falls into each category for each city, with darker red indicating higher concentrations of the given label. Some notable insights are that \textit{Racist} content remains consistently low across all cities (<0.3\%), and \textit{Harmful Generalization} and \textit{Deserving}/\textit{Undeserving} samples occur more frequently in data sourced from cities such as Kalamazoo and San Francisco than in cities like Fayetteville and South Bend. Geographic variation in content patterns suggests city-specific discourse characteristics potentially influenced by local political climates, demographics, and platform usage patterns.

Our analysis does not reveal a significant difference between small and large cities in terms of bias against PEH when grouping all of the datasources together.  Furthermore, there is no significant difference temporally. Therefore, the data must be split by source in order to reveal details as highlighted in Figure \ref{fig:bias_city_size} (certain data sources are more prevalent for different cities).

\section{LLM Classification Results}
\label{sec:llm_results}
Main-text Table~\ref{tab:llm_benchmark} summarizes macro-F1 on the 1,698-item gold-standard set (2-of-3 majority; macro F1 = mean per-label F1). Table~\ref{tab:detailed_f1_scores_soft} gives the full zero- and few-shot grid with micro-F1 for all six prompt LLMs.

\subsection{Soft-Label F1: Prompt LLMs (Zero- and Few-Shot)}

\begin{table*}[htpb!]
\resizebox{\textwidth}{!}{
\begin{tabular}{lcccccccccccccccccc}
\toprule
Data Source & \multicolumn{2}{c}{LLaMA} & \multicolumn{2}{c}{Phi-4} & \multicolumn{2}{c}{Qwen} & \multicolumn{2}{c}{Gemini} & \multicolumn{2}{c}{Grok} & \multicolumn{2}{c}{GPT} \\
& Zero & Few & Zero & Few & Zero & Few & Zero & Few & Zero & Few & Zero & Few \\
\midrule
Reddit (Macro) & 19.42 & 29.66 & 15.33 & 24.09 & 35.62 & 38.24 & 50.90 & \textbf{51.52} & 38.51 & 38.61 & 50.57 & 51.28 \\
Reddit (Micro) & 41.41 & 40.71 & 32.06 & 39.38 & 49.80 & 53.99 & 65.18 & \textbf{66.36} & 47.94 & 50.80 & 63.13 & 65.22 \\
X (Twitter) (Macro) & 16.57 & 25.50 & 11.90 & 26.60 & 28.00 & 35.45 & 45.06 & \textbf{45.91} & 26.57 & 35.37 & 43.21 & 44.65 \\
X (Twitter) (Micro) & 39.56 & 39.32 & 21.84 & 44.07 & 39.45 & 54.99 & 66.30 & \textbf{67.39} & 46.70 & 54.00 & 60.39 & 63.59 \\
News (Macro) & 15.05 & 15.94 & 17.11 & 21.30 & 22.62 & 25.12 & 35.99 & \textbf{39.08} & 26.04 & 28.69 & 34.79 & 37.93 \\
News (Micro) & 36.57 & 28.33 & 26.85 & 50.13 & 35.31 & 55.25 & 65.03 & \textbf{68.81} & 51.66 & 54.10 & 61.20 & 64.26 \\
Meeting Minutes (Macro) & 13.84 & 21.54 & 18.12 & 20.81 & 26.82 & 30.32 & 38.06 & \textbf{41.38} & 29.46 & 34.46 & 37.70 & 39.96 \\
Meeting Minutes (Micro) & 39.97 & 38.94 & 29.31 & 39.55 & 46.09 & 60.04 & 66.64 & \textbf{71.12} & 48.63 & 62.39 & 61.13 & 67.89 \\
Family Avg (Macro) & \multicolumn{2}{c}{19.69} & \multicolumn{2}{c}{19.41} & \multicolumn{2}{c}{30.27} & \multicolumn{2}{c}{\textbf{43.49}} & \multicolumn{2}{c}{32.21} & \multicolumn{2}{c}{42.51} \\
Family Avg (Micro) & \multicolumn{2}{c}{38.10} & \multicolumn{2}{c}{35.40} & \multicolumn{2}{c}{49.36} & \multicolumn{2}{c}{\textbf{67.11}} & \multicolumn{2}{c}{52.03} & \multicolumn{2}{c}{63.35} \\
\bottomrule
\end{tabular}
}
\centering
\caption{Soft-label macro and micro F1 for all prompt LLMs by data source (gold-standard set; 2-of-3 majority reference; macro F1 = mean per-label F1).}
\label{tab:detailed_f1_scores_soft}
\end{table*}

Gemini attains the highest family-average macro/micro F1; GPT-4.1 is competitive on Reddit. Few-shot helps marginally for several closed APIs but does not remove prevalence-gap miscalibration (Section~\ref{sec:ro3_calibration}).

\section{LLM-as-Classifier vs.\ Pseudo-Label Validation}
\label{sec:evaluation_framing}
The main text distinguishes two ML roles (Sections~\ref{sec:llm_eval} and~\ref{sec:ro3_calibration}).

\subsection{Model Training, Splits, and LoRA}
\label{sec:model_training_splits}
We benchmark six prompt LLMs (GPT-4.1, Gemini 2.5 Pro, Grok-4, LLaMA 3.2 3B, Qwen 2.5 7B, Phi-4 Mini) against transformer encoders (BERT, RoBERTa, ModernBERT). Closed APIs are evaluated zero- and few-shot on the 1,698-item gold-standard set (2-of-3 majority reference). Local LLMs are first evaluated zero-shot on the same reference set; they are then fine-tuned with LoRA \citep{hu2022lora} on GPT-4.1 pseudo-labels for the unlabeled corpus ($\approx$48,389 items). LoRA adapts pretrained weights $W_0$ via low-rank updates:
\begin{equation}
W' = W_0 + BA, \quad B \in \mathbb{R}^{d \times r}, A \in \mathbb{R}^{r \times k}
\end{equation}
with $r \ll \min(d, k)$ so only a small fraction of parameters is trainable.

\textbf{Splits.} Prompt and LoRA-on-GPT models are never scored against teacher labels on a held-out test split: all reported F1 uses gold-standard reference labels only. For encoders trained on GPT pseudo-labels, training uses the full pseudo-labeled corpus; the gold-standard audit set is split 50/50 (validation/test, \texttt{random\_state=42}) to tune per-label decision thresholds and report macro/micro F1. Gold-only encoder runs use a 70/10/20 train/validation/test split per source on reference annotations. Given severe class imbalance (e.g., ``racist'' in $<$1\% of samples), encoders select thresholds on validation:
\begin{equation}
\tau^* = \argmax_{\tau \in [0,1]} F_1^{\text{macro}}(y_{\text{val}}, \hat{y}_{\tau})
\end{equation}
and report scores on the held-out test set. Training uses objectives suited to extreme imbalance (focal loss and threshold tuning). Table~\ref{tab:detailed_f1_scores_transformers} labels local LLM finetuning as LoRA-on-GPT and encoder columns as gold-only vs.\ GPT-pseudo training.

\subsection{Gold vs.\ GPT Pseudo-Labels (All Models)}

\begin{table*}[!t]
\centering
\resizebox{\textwidth}{!}{
\begin{tabular}{lccccccccccccccc}
\toprule
Data Source & \multicolumn{2}{c}{LLaMA} & \multicolumn{2}{c}{Phi-4} & \multicolumn{2}{c}{Qwen} & Gemini & Grok & GPT & \multicolumn{2}{c}{BERT} & \multicolumn{2}{c}{RoBERTa} & \multicolumn{2}{c}{ModernBERT} \\
 & Zero & Finetuned & Zero & Finetuned & Zero & Finetuned & Zero & Zero & Zero & Gold & Pseudo & Gold & Pseudo & Gold & Pseudo \\
\midrule
Reddit (Macro) & 19.42 & 44.18 & 15.33 & 33.87 & 35.62 & 50.13 & \textbf{50.90} & 38.51 & 50.57 & 25.45 & 28.20 & 27.22 & 24.17 & 32.11 & 49.56 \\
Reddit (Micro) & 41.41 & 56.70 & 32.06 & 52.51 & 49.80 & 61.66 & \textbf{65.18} & 47.94 & 63.13 & 42.50 & 41.35 & 38.13 & 45.30 & 41.73 & 64.42 \\
News (Macro) & 15.05 & 15.67 & 17.11 & 15.67 & 22.62 & 15.22 & \textbf{35.99} & 26.04 & 34.79 & 11.08 & 16.50 & 14.24 & 14.27 & 16.18 & 23.63 \\
News (Micro) & 36.57 & 47.72 & 26.85 & 47.72 & 35.31 & 53.42 & 65.03 & 51.66 & 61.20 & 33.77 & 45.71 & 45.02 & 50.93 & 33.33 & \textbf{69.27} \\
Meeting Minutes (Macro) & 13.84 & 14.93 & 18.12 & 19.58 & 26.82 & 21.65 & \textbf{38.06} & 29.46 & 37.70 & 11.66 & 19.16 & 15.64 & 18.37 & 22.82 & 33.86 \\
Meeting Minutes (Micro) & 39.97 & \textbf{72.76} & 29.31 & 57.42 & 46.09 & 67.00 & 66.64 & 48.63 & 61.13 & 37.86 & 57.61 & 39.58 & 43.61 & 48.87 & 72.54 \\
X (Twitter) (Macro) & 16.57 & 20.99 & 11.90 & 20.99 & 28.00 & 19.98 & \textbf{45.06} & 26.57 & 43.21 & 18.07 & 16.28 & 18.17 & 16.25 & 30.75 & 33.88 \\
X (Twitter) (Micro) & 39.56 & 42.76 & 21.84 & 42.76 & 39.45 & 48.78 & \textbf{66.30} & 46.70 & 60.39 & 26.14 & 59.45 & 47.44 & 53.74 & 43.82 & 64.39 \\
Avg (Macro) & 16.22 & 23.94 & 15.62 & 22.53 & 28.27 & 26.75 & \textbf{42.50} & 30.14 & 41.57 & 16.56 & 20.04 & 18.82 & 18.27 & 25.46 & 35.23 \\
Avg (Micro) & 39.38 & 54.98 & 27.52 & 50.10 & 42.66 & 57.71 & \textbf{65.79} & 48.73 & 61.46 & 35.07 & 51.03 & 42.54 & 48.39 & 41.94 & 67.66 \\
\bottomrule
\end{tabular}
}
\caption{Macro/micro F1 by source: prompt LLMs (zero-shot / LoRA-on-GPT) and encoders (gold vs.\ GPT pseudo-labels). Finetuned = LoRA (local LLMs) or val-opt thresholds (encoders).}
\label{tab:detailed_f1_scores_transformers}
\end{table*}

\paragraph{Primary: LLMs as classifiers on the gold standard.}
Table~\ref{tab:detailed_f1_scores_soft} benchmarks six prompt LLMs on gold-standard soft labels. Corpus-wide exploratory prevalence (Appendix~\ref{sec:exploratory_corpus}) uses GPT-4.1 few-shot labels on $\approx$50k texts; Section~\ref{sec:ro3_calibration} audits agreement on the gold-standard set only.

\paragraph{Secondary: pseudo-labels validate scalability, not truth.}
GPT pseudo-labels on 48,389 additional items ask whether encoder distillation is \emph{feasible} when gold is too small to train, not whether teacher prevalence equals gold-standard reference prevalence. Table~\ref{tab:detailed_f1_scores_transformers} compares prompt zero-shot / LoRA-on-GPT scores with BERT, RoBERTa, and ModernBERT trained on gold only vs.\ on GPT pseudo-labels.

\paragraph{NIMBY prevalence gap after pseudo-label training.}
Encoders with val-opt thresholds reach $\approx 0$\,pp NIMBY gap on the gold test split (Table~\ref{tab:nimby_pseudolabel_mitigation}); LoRA local LLMs stay within $\pm 3$\,pp vs.\ gold reference, not teacher prevalence.

\begin{table}[!htb]
\centering
\small
\begin{tabular}{@{}llr@{}}
\toprule
\textbf{Family} & \textbf{Setting} & \textbf{Gap (pp)} \\
\midrule
\multicolumn{3}{@{}l@{}}{\textit{Fine-tuned}} \\
 & BERT (val-opt) & $-1.1$ \\
 & RoBERTa (val-opt) & $+0.0$ \\
 & ModernBERT (val-opt) & $+0.0$ \\
 & LLaMA LoRA (val-opt) & $+3.0$ \\
 & Phi-4 LoRA (val-opt) & $+2.4$ \\
 & Qwen LoRA (val-opt) & $-1.9$ \\
\bottomrule
\end{tabular}
\caption{NIMBY gap after GPT pseudo-label training (val-opt; gold test split). Encoders $\approx 0$\,pp; LoRA LLMs $\pm 3$\,pp.}
\label{tab:nimby_pseudolabel_mitigation}
\end{table}

\paragraph{Why this validates a theory but is not the deployment solution.}
Gold-only encoders stay weak (ModernBERT 25.46 avg macro-F1) while GPT-pseudo-label training raises ModernBERT to 35.23, showing that \textbf{scaling labels beyond the audit set can work}. That does not make pseudo-labels ground truth: teacher drift and prevalence gaps mean partners must keep stratified gold-standard audits before acting on corpus dashboards. Distillation inherits teacher bias; it is a feasibility check, not a substitute for human judgment.

GPT-4.1 few-shot is the standard teacher for exploratory corpus labeling and pseudo-label generation; Gemini 2.5 Pro is within $\approx$1 macro-F1 point on gold (Appendix~\ref{sec:teacher_model}). Teacher labels are validation inputs, not municipal ground truth.

\section{OATH Flan-T5 Transfer Baseline}
\label{sec:oath_baseline}
OATH-Frames \citep{ranjit2024oath} trains Flan-T5-Large as a multilabel scaler and reports aggregated macro F1 $\approx$50--51 on X---i.e., the mean of per-frame F1 over their nine issue-specific frames (plus a Public Opinion filter in their table), (NIMBY F1 $=26$ on $|\mathcal{D}_{\mathrm{test}}|{=}1280$). We evaluate that released tagger on our 1,698-item gold-standard set for the same nine issue-specific frames, using their published training prompt prefix (comma-separated snake\_case labels). Our macro-F1 is the unweighted mean of those nine per-frame F1s under 2-of-3 soft-label reference. Table~\ref{tab:oath_vs_prompted} compares nine-frame macro-F1 against our prompted LLMs; Table~\ref{tab:oath_gold_frame_eval} gives per-frame precision, recall, F1, and prevalence gap.

On our multi-domain gold the scaler reaches nine-frame macro-F1 $42.1$ (NIMBY F1 $17.5$), matching few-shot Gemini/GPT-4.1 on the overlapping subset and beating local Qwen ($31.7$) and LLaMA/Phi-4 ($\approx$22--23), but $\approx$8 points below its in-domain X Aggregated Macro. Mean prevalence gap is $+3.8$\,pp: solutions/interventions is strongly over-tagged ($+23.3$\,pp) while money/aid and societal critique are slightly under-tagged. Unlike many prompted LLMs' large NIMBY inflation (Section~\ref{sec:ro3_calibration}), OATH's NIMBY gap is modest ($+1.8$\,pp) but F1 remains low. Relative to our GPT-pseudo-label encoders (Appendix~\ref{sec:evaluation_framing}), OATH is stronger on nine-frame F1 than ModernBERT ($\approx$35 avg macro over all 16 labels; $\approx$31 on the nine overlapping frames) and LoRA-tuned local LLMs ($\approx$23--27), but those fine-tunes cover the full 16-category taxonomy and can compress NIMBY prevalence gaps to $\approx$0--2\,pp with validation-tuned thresholds, and the OATH scaler is not calibrated for this. This is a transfer check with the correct inference prefix, not a re-score of OATH's original expert-annotated X test set.

\begin{table}[!htb]
\centering
\scriptsize
\setlength{\tabcolsep}{3pt}
\begin{tabular}{@{}lcc@{}}
\toprule
\textbf{Model} & \textbf{Setting} & \textbf{Macro-F1} \\
\midrule
OATH Flan-T5-Large & released scaler & \textbf{42.10} \\
Gemini 2.5 Pro & few-shot & \textbf{42.40} \\
GPT-4.1 & few-shot & \textbf{42.10} \\
Qwen 2.5 7B & few-shot & 31.70 \\
Phi-4 Mini & few-shot & 23.30 \\
LLaMA 3.2 3B & few-shot & 22.00 \\
\bottomrule
\end{tabular}
\caption{Nine-frame macro-F1 on our pooled gold (2-of-3). OATH in-domain Aggregated Macro on X is $\approx$50 (NIMBY alone 26).}
\label{tab:oath_vs_prompted}
\end{table}

\begin{table}[!htb]
\centering
\scriptsize
\setlength{\tabcolsep}{2pt}
\begin{tabular}{@{}lrrrrr@{}}
\toprule
\textbf{Frame} & $n_{+}$ & P & R & F1 & Gap \\
\midrule
MoneyAid & 309 & 76.31 & 61.49 & 68.10 & $-3.5$ \\
GovCrit & 203 & 41.85 & 75.86 & 53.94 & $+9.7$ \\
SocCrit & 152 & 38.61 & 25.66 & 30.83 & $-3.0$ \\
SolnInt & 678 & 56.76 & 89.82 & 69.56 & $+23.3$ \\
Interact & 80 & 35.71 & 62.50 & 45.45 & $+3.5$ \\
MediaPort & 10 & 17.39 & 40.00 & 24.24 & $+0.8$ \\
NIMBY & 42 & 13.89 & 23.81 & 17.54 & $+1.8$ \\
HarmGen & 118 & 43.88 & 51.69 & 47.47 & $+1.2$ \\
(Un)Deserv & 23 & 18.75 & 26.09 & 21.82 & $+0.5$ \\
\midrule
MACRO & --- & 38.13 & 50.77 & 42.11 & $+3.8$ \\
\bottomrule
\end{tabular}
\caption{Released OATH Flan-T5-Large scaler vs.\ gold 2-of-3 on nine overlapping frames ($n{=}1{,}698$)..}
\label{tab:oath_gold_frame_eval}
\end{table}

\end{document}